\documentclass[twocolumn]{aastex701}

\begin{document}

\title{TESS Photometry and Radial Velocity Analysis of the sub-Neptune Exoplanet $\pi$~Mensae~c and the Wider $\pi$~Mensae Planetary System}

\author[orcid=0009-0009-6718-6595]{Skylar S. Larsen}
\affiliation{Department of Astronomy and Carl Sagan Institute, Cornell University, Ithaca, New York 14853, USA}
\email[show]{ssl248@cornell.edu}

\author[orcid=0000-0001-7836-1787]{Jake D. Turner}
\affil{Department of Astronomy and Carl Sagan Institute, Cornell University, Ithaca, New York 14853, USA}
\email[show]{jaketurner@cornell.edu}

\author[orcid=0000-0002-8507-1304]{Nikole K. Lewis}
\affiliation{Department of Astronomy and Carl Sagan Institute, Cornell University, Ithaca, New York 14853, USA}
\email{nikole.lewis@cornell.edu}

\author[orcid=0000-0002-4416-8011]{Maryame El~Moutamid}
\affiliation{Southwest Research Institute, 1301 Walnut St Suite 400, Boulder, CO 80302}
\email{maryame.elmoutamid@swri.edu} 

\author[0000-0003-0360-4451]{Diana Solano-Oropeza}
\affiliation{Department of Astronomy and Carl Sagan Institute, Cornell University, Ithaca, New York 14853, USA}
\email{ds2396@cornell.edu}

\author[0000-0001-8703-7751]{Lili Alderson} 
\affiliation{Department of Astronomy, Cornell University, 122 Sciences Drive, Ithaca, NY 14853, USA}
\email{lili.alderson@cornell.edu}

\author[orcid=0000-0002-4945-1860]{V. Abby Boehm}
\affiliation{Department of Astronomy and Carl Sagan Institute, Cornell University, Ithaca, New York 14853, USA}
\email{vab55@cornell.edu}

\author[0000-0003-4755-584X]{Daniel A. Yahalomi}
\affiliation{Kavli Institute for Astrophysics and Space Research, Massachusetts Institute of Technology, Cambridge, MA 02139, USA}
\email{yahalomi@mit.edu}

\author[0000-0001-9796-2158]{Emily K.~Deibert}
\affiliation{Department of Physics and Astronomy, University of Waterloo, 200 University Avenue West, Waterloo, Ontario N2L 3G1, Canada}
\affiliation{Waterloo Centre for Astrophysics, University of Waterloo, Waterloo, Ontario N2L 3G1, Canada}
\email{emily.deibert@uwaterloo.ca}

\author[0000-0001-5349-6853]{Ray Jayawardhana}
\affiliation{Department of Astronomy, California Institute of Technology, Pasadena, CA 91125, USA}
\email{rayjay@caltech.edu}

\author[0000-0002-1934-6250]{Dong Lai}
\affiliation{Center for Astrophysics and Planetary Science, Department of Astronomy, Cornell University, Ithaca, NY 14853, USA}
\affiliation{Tsung-Dao Lee Institute, Shanghai Jiao Tong University, 1 Lisuo Road, Shanghai 201210, China}
\email{dl57@cornell.edu}

\author[0000-0002-4451-1705]{Adam B. Langeveld}
\affiliation{Department of Physics and Astronomy, Johns Hopkins University, Baltimore, MD 21218, USA}
\affiliation{Department of Astronomy and Carl Sagan Institute, Cornell University, Ithaca, New York 14853, USA}
\email{abl224@cornell.edu}

\author[0000-0002-5785-9073]{Kyle A. Pearson}
\affiliation{Deep Learning Analytics, General Dynamics Mission Systems, Scottsdale, AZ 85257, USA }
\email{kyle.pearson@gd-ms.com}

\begin{abstract}
Exoplanet characterization relies on precise measurements of planetary orbital and physical parameters. This is particularly important for planetary dynamics and atmospheric evolution, as orbital parameters help constrain system evolution, resolve ambiguities, and gauge atmospheric retention. The first exoplanet discovered by the Transiting Exoplanet Survey Satellite (TESS), $\pi$~Men~c, is a warm sub-Neptune orbiting a bright Sun-like star in a system containing (at least) one other planet with a wildly different period and size. Lying near the 1.5-2.0 $R_{\oplus}$ radius gap, $\pi$~Men~c is expected to have lost its primordial hydrogen and helium, but kept heavier compounds like H\textsubscript{2}O and CO\textsubscript{2}. The $\pi$ Men system is well observed with decades of radial velocity measurements, and TESS has continued to observe $\pi$~Men~c, yielding six years and 21 sectors of photometry. We present a comprehensive analysis of these TESS data and 22 years of radial velocity measurements to provide updated orbital ephemerides for $\pi$~Men~b, c, and the proposed third planet, $\pi$~Men~d. Our newly derived $\pi$~Men~c period error margins are an order of magnitude improved from previous estimates, and we estimate the mass range of $\pi$~Men~d to be 13.4 $\leq$ M$_d$ $<$ 20 M$_{\oplus}$. We find that $\pi$~Men~c is a uniquely interesting target for future transmission spectroscopy studies with JWST, and that existing radial velocity data are consistent with the existence of a third planet.
\end{abstract}

\keywords{}
 
\section{Introduction} \label{sec:intro}

Rates of exoplanet detection and characterization have exploded in recent decades, with over 6,000 exoplanets discovered and confirmed at the time of writing\footnote{NASA Exoplanet Archive: \citep{Christiansen2025}}. These exoplanets show great diversity in size, orbital parameters, and equilibrium temperature, with the majority having no Solar System counterparts. The indirect methods by which these planets are discovered, including the transit method \citep{Charbonneau2000, Henry_2000, Cameron2016}, radial velocity method \citep{Mayor_1995}, and transit timing variations (TTVs; \citealt{Agol_2005, Holman_2005, Nesvorny_2012}) have been crucial to uncovering this wider diversity, and are often used in tandem whenever possible to return thorough analyses of single systems. Orbital and planetary characterizations are capable of constraining the exoplanet's atmospheric climate and dynamical history \citep{Kaspi_2015, Zahnle_Catling_2017}; parameters such as a planet's radius, mass, and orbital period are crucial for understanding its equilibrium temperature, escape velocity, and instellation (among other parameters), values which are important for understanding the planet's atmospheric conditions and dynamical past.


The $\pi$ Mensae system, consisting of a bright (J~=~4.87, V~=~5.67), G-type Sun-like star, $\pi$ Men A \citep{Cutri_2003}, and at least two planets, is of particular interest for dynamical, atmospheric, and observational reasons. The first-discovered and outermost planet, $\pi$~Men~b ($P \simeq 2088.33$ days, $M_p \simeq 12.6 M_{J}$, $e \simeq 0.64$; \citealt{Hatzes_2022}), is a highly eccentric, near brown dwarf-mass object discovered in 2002 via radial velocity measurements \citep{Jones_2002}. The second-discovered and innermost planet, sub-Neptune exoplanet $\pi$~Men~c ($P \sim$ 6.27 days, $M_p \sim$  3.63 M$_{\oplus}$; \citealt{Hatzes_2022}), was discovered in 2018 (\citealt{Huang_2018,Gandolfi_2018}) and was the first exoplanet discovered by NASA's Transiting Exoplanet Survey Satellite (TESS; \citealt{Ricker2015}). With a radius of 2.145 ± 0.015 R$_{\oplus}$ and mass of 3.63±0.38 M$_{\oplus}$, $\pi$~Men~c has a bulk density of 2 g/cm$^3$ and is consistent with a gaseous composition (\citealt{Hatzes_2022}) according to the definitions set by \citet{Zeng_2016}. In part due to $\pi$~Men~b's early discovery date and easily detectable radial velocity amplitude, the $\pi$ Men system has decades of radial velocity measurements available for study, and new discoveries regarding the system are still being made. A tentative radial velocity detection of a third planet in the system, $\pi$~Men~d, was posited 20 years after the initial detection of the first planet \citep{Hatzes_2022, Laliotis_2023, Harada_2025}. 


The innermost planet, sub-Neptune $\pi$~Men~c, is of particular interest for exoplanet atmospheric science. The planetary radius gap is an apparent deficiency of detected close-in exoplanets between 1.5–2.0 R$_{\oplus}$; the atmospheres of these planets are thought to be actively escaping (\citealt{Owen_Wu_2013,Fulton_2017}). $\pi$~Men~c exists on the edge of the planetary radius gap \citep{Owen_Wu_2013, Fulton_2017} and may be losing its atmosphere \citep{King_2019}, a conclusion supported by the observation of escaping C II ions \citep{García_Muñoz_2021}. While \citet{King_2019} predicted that $\pi$~Men~c would be bright enough in the Ly $\alpha$ wavelength region for a hydrogen atmospheric escape detection, \cite{García_Muñoz_2020} reported a Ly $\alpha$ nondetection, supporting the hypothesis that $\pi$~Men~c is not entirely hydrogen-dominated. Some predict that $\pi$~Men~c’s atmosphere could be at least half heavy volatiles by mass \citep{García_Muñoz_2021} and could have the highest fraction of heavy atmospheric volatiles in an exoplanet to date. Its instellation and escape velocity predict that it will retain heavy molecules such as H\textsubscript{2}O and CO\textsubscript{2}, but lose lighter primary atmospheric gases. As an older planet (age = 3.92 Gyr) \citep{Hatzes_2022} with a potentially escaping atmosphere \citep{García_Muñoz_2021}, $\pi$~Men~c appears contradictory. Despite its unique properties, little is known about $\pi$~Men~c’s atmosphere except for the confirmed presence of escaping C II ions \citep{García_Muñoz_2021} and a Ly $\alpha$ nondetection \citep{García_Muñoz_2020}. As a super-Earth/sub-Neptune on the edge of the planetary radius gap with a potentially heavy atmosphere orbiting a Sun-like star, $\pi$~Men~c is fascinating from a planetary evolution perspective: it has no solar system equivalent, and the extent of its atmospheric escape will constrain its evolution. And as a planet in a keystone radial velocity system, with years of space-based transit photometry data and decades of radial velocity measurements, $\pi$~Men~c provides a unique opportunity to help inform the orbital and planetary parameters for all planets in the system.


In this paper, we combine six years' worth of TESS transit photometry data and over twenty years of archival radial velocity measurements from various sources into a comprehensive analysis to provide updated orbital and planetary parameters for $\pi$~Men~c, as well as updated radial velocity parameters for the other planet(s) in the system. Our goals are to improve the $\pi$ Men planets' parameters for the purposes of further observation and modeling, to better constrain $\pi$~Men~c's ability to retain its atmospheric components, and to investigate the possibility of the existence of $\pi$~Men~d.

This paper is organized as follows. Section \ref{sec:Phot_Model} details our photometry and transit timing variation observations and data sources (Section \ref{subsec:Photometry_Observe}), data analysis (Section \ref{subsec:TTV_DataAnalyis}), and results (Section \ref{subsec:TTV_Results}). Section \ref{sec:RV} summarizes our radial velocity model observations and data sources (Section \ref{subsec:RV_Observe}), data analysis (Section \ref{subsec:RV_DataAnalyis}), and results (Section \ref{subsec:RV_Results}). Section \ref{sec:discussion} is split into two parts: Section \ref{subsec:disc_1} delves into the atmospheric implications of $\pi$~Men~c's newly updated parameters, Section \ref{subsec:disc_2} explores the radial velocity results' implications for the existence of a third $\pi$~Men planet, and Section \ref{subsec:disc_3} explores the orbital dynamics of a three-planet $\pi$~Men system. We summarize our conclusions in Section \ref{sec:conclusion}.


\section{TESS PHOTOMETRY} \label{sec:Phot_Model}

\subsection{Photometry Observations} \label{subsec:Photometry_Observe}

$\pi$~Men~c (TOI-144.01), being the first exoplanet discovered by TESS, was first observed in TESS Sector 1 (July 25, 2018 to August 22, 2018). Table \ref{tab:planet_parameters} lists stellar and planetary parameters derived by previous research. Since its discovery, $\pi$~Men~c has been observed by a total of 21 sectors (doi:\href{https://archive.stsci.edu/doi/resolve/resolve.html?doi=10.17909/t9-yjj5-4t42}{10.17909/t9-yjj5-4t42}), and was most recently observed by Sector 95 (July 25, 2025 to August 20, 2025). These observations were processed by the Science Processing Operations Center (SPOC) pipeline, which produces light curves corrected for systematics and searches for transiting planets \citep{Jenkins_2016}. All of the data products produced by SPOC are publicly available from the Mikulski Archive for Space Telescopes (MAST)\footnote{All of the SPOC data products are publicly accessible from the Mikulski Archive for Space Telescopes at \url{https://archive.stsci.edu/}}. We downloaded light curve (LC) files as data products via the software program \texttt{lightkurve} \citep{lightkurve_2018}, with all light curves having exposure times of 120 seconds. The Presearch Data Conditioning (PDC) component of the SPOC pipeline corrects the light curves for pointing or focus-related instrumental signatures and discontinuities resulting from radiation events in the CCD detectors, outliers, and flux contamination \citep{Jenkins_2016}. The light curve resulting from the PDC corrections is recorded as the PDCSAP{\_}FLUX; this light curve was the data product considered in our analysis. As shown in our previous studies \citep{RiddenHarper2020,Turner2021,Turner2022}, this light curve product is adequate to study TTVs. The PDCSAP{\_}FLUX is further processed by using a running median filter to remove any long-period systematics before the SPOC pipeline searches for transits.  

\begin{table}
\caption{Physical and Orbital Parameters of $\pi$~Men~c and Stellar Parameters of $\pi$ Men as Reported by Previous Research}
    \centering
    \begin{tabular}{lccc}
     \hline
     Parameter   & value  & Source \\
     \hline
  P [Days] & 6.267852$\pm$0.000016  &  [3] \\
  $R_{p}/R_{*}$ & 0.0165$\pm$0.0001 & [1]\\
  $R_{p}$ [$R_{J}]$ & 0.188$\pm$0.004 & [1] \\
  $a/R_{*}$ & 13.38$\pm$0.26 &     [2]  \\
  a [AU] & 0.069$\pm$0.003 & [5]                   \\
  i [\textdegree] & 87.05$\pm$0.15 & [3] \\
  T$_{dur}$ [hrs] & 2.969$^{+0.030}_{-0.032}$  & [4] \\
  $M_{p}$ [$M_{J}]$ & 0.0114$\pm$0.0012 & [3] \\
  $T_{eq}$ [K] & 1169.8$^{+2.8}_{-4.3}$ & [2]\\
   \hline
  $R_*$ [$R_\odot$]  & 1.190 $\pm$ 0.004   & [3] \\
  $M_*$ [$M_\odot$] & 1.07 $\pm$ 0.04   & [3]\\
  $L_*$ [$L_\odot$] & 1.57 $\pm$ 1.00   & [6] \\
  $T_{eff*}$ [K]&  5998$\pm$62  & [3] \\
  TESS Mag & 5.1054$\pm$0.006 & [7]\\
   \hline
    \end{tabular}
    \label{tab:planet_parameters}
    \tablecomments{The stellar parameters $R_*$, $M_*$, $L_*$, and $T_{eff*}$ were only used to calculate planetary mass, equilibrium temperature, instellation, and transmission spectroscopy metric (TSM) as described in Section \ref{sec:discussion}. 
    }
    \tablerefs{
    [1] \citet{Damasso_2020}; 
    [2] \citet{Huang_2018}; 
    [3] \citet{Hatzes_2022};
    [4] \citet{Gandolfi_2018}; 
    [5] \citet{Feng_2022};
    [6] \citet{GAIADR3_2023}; 
    [7] \citet{Stassun2018}
    }
\end{table}

Of the twenty-one available $\pi$~Men~c sectors, only one sector (Sector 27) was rendered unusable by a severe influx of scattered light \citep{Sector27} and discarded. Portions of Sectors 4 and 8, namely the regions between 1420-1424 and 1535-1536 BTJD (TESS Barycentric Julian Day, BTJD = BJD - 2457000), were also obscured by severe scattered light \citep{Sector4,Sector8} and discarded; these regions did not contain transits. The first transit of Sector 4 and the second transit of Sector 68 were excluded from data analysis, as the transits were obscured by severe stellar fluctuations. A full list of photometry sectors is available in the Appendix (Table \ref{tab:Phot_sectors}).

\subsection{Photometry Data Analysis} \label{subsec:TTV_DataAnalyis}

Using the python modeling tool \texttt{juliet} \citep{Espinoza_2019}, we model all 21 sectors of TESS photometry data simultaneously to derive orbital/planetary parameters and detect possible Transit Timing Variations (TTVs), while at the same time detrending systematics via a time-series Gaussian Process (GP). Either via direct nested sampling or further derivation, we calculate estimates of $\pi$~Men~c's orbital and planetary parameters: scaled semi-major axis ($a/R^*$), stellar density ($\rho_*$) in $\text{kg}/\text{m}^3$, transit midpoint ($t_0$) in TESS Barycentric Julian Date (BTJD), inclination ($i$) in degrees, two photometric limb darkening parameters ($q_1$ and $q_2$), each and every transit midpoint time designated by the function $T(n)$, and two free parameters ($r_1$ and $r_2$) which are used to further calculate the planet-to-star radius ratio ($R_p / R^*$) and impact parameter ($b$) as described in \citet{Espinoza_2018}. \texttt{juliet} makes use of the python package \texttt{batman} \citep{batman} for modeling exoplanet transit light curves. 

The free parameters $r_1$ and $r_2$, stellar density $\rho_*$, limb-darkening parameters $q_1$ and $q_2$, and transit midpoints $t_n$ were provided with flexible priors and solved for via nested sampling. We used uniform priors of $r_1$ and $r_2$ to explore every physical combination of both parameters. Stellar density $\rho_*$ was provided with a wide loguniform prior informed by previous estimates \citep{Huang_2018, Damasso_2020}. 

\texttt{juliet} solves for the limb-darkening parameters via the parameterized quadratic limb darkening law described by \citet{Kipping_2013}, in which the two parameters $q_1$ and $q_2$ are sampled between the range between 0 and 1, assuming each set of sampled values give rise to physical solutions \citep{Espinoza_2019}. The initial values for the limb darkening parameters were calculated via the python package \texttt{ExoTiC-LD} \citep{Grant_2024} using the stellar parameters provided by \citet{Damasso_2020}, according to the parameterization described by \citet{Kipping_2013}. The limb darkening parameters were set to a normal distribution with a wide parameter space (Table \ref{tab:photom_run_numbers}) to encourage ample exploration of limb darkening parameter options.

\begin{figure*}
  \begin{center}
    \includegraphics[width=0.47\textwidth]{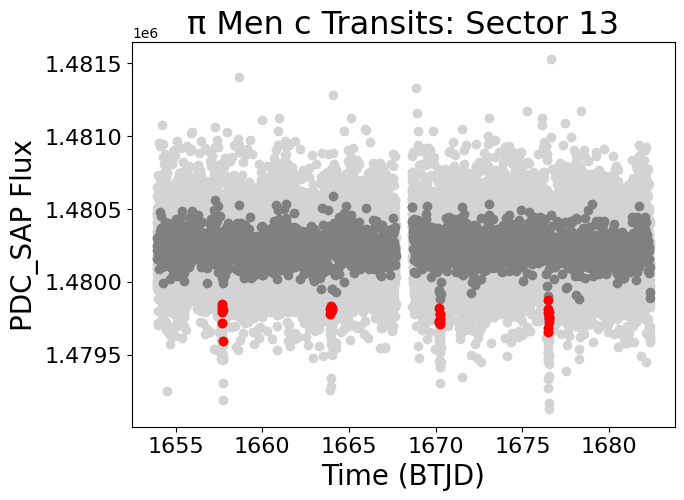}
    \includegraphics[width=0.50\textwidth]{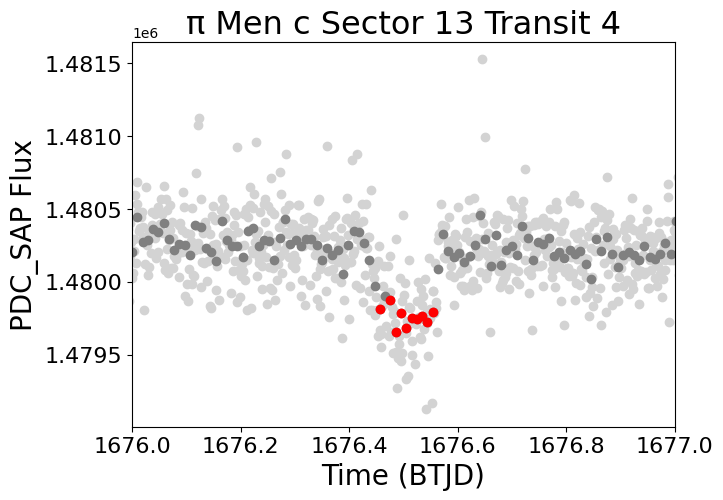}
    \caption{The process of deducing transit midpoint priors from $\pi$~Men~c TESS photometry data. The x-axis is listed in units of TESS Barycentric Julian Date (BTJD), which is equal to BJD - 2457000. Top: TESS sector 13. The dark grey points are binned, and light grey unbinned. All binned data points the pipeline flags as being part of a transit ($Z \leq -3.3$ in this case) are highlighted in red. Bottom:  A singular transit. The transit midpoint initial guess is the median of the red points. This process allows for an automatic and streamlined selection of transit midpoint priors.
    }
    \label{fig:inator}
  \end{center}
\end{figure*}

We chose the transit midpoint priors via the following automatic process, as detailed in Figure \ref{fig:inator}. $\pi$~Men~c is a sub-Neptune orbiting a G-type star, and as such systematically differentiating each transit from the surrounding noise is non-trivial. First, each sector of TESS photometry time-series data was individually binned to a cadence of 840 seconds (7 data points per bin), to smooth over stellar noise and instrumental artifacts and make the transits more distinguishable from scatter. We then identified any points in the binned data below a certain Z-score that would isolate only data points within a transit using \texttt{scipy.stats.zscore} \citep{scipy_2020}. The median of each cluster of binned points identified as a single transit became the transit midpoint prior ($t_n$). The Z-score required to isolate transit data points and exclude baseline and instrumental/stellar features varied by sector, but was reliably $-4 \leq Z \leq -3$. In the nested sampling model itself, each prior midpoint was sampled via a normal distribution wide enough to consider the parameter space within the transit duration and largely exclude the space outside of it.

The eccentricity ($e$) and argument of periastron ($\omega$) were kept fixed to 0 and 90\textdegree respectively in the photometry model to reduce computation time. The argument of periastron has never been well constrained by previous research (especially without complementary radial velocity data), and previous studies either assumed an eccentricity of zero for $\pi$~Men~c or found the eccentricity to be near-zero with reasonable precision.

Other orbital and planetary parameters ($P$, $t_0$, $a/R*$, $b$, and $i$), are computed by \texttt{juliet} in ways other than via nested sampling. To solve for TTVs, instead of solving for the orbital period or initial mid-transit time via direct nested sampling, we solved for each transit midpoint individually ($t_n$) and derived the period and transit midpoint from a least-squared best-fit line of all transit midpoints ($T[n]$), with the period ($P$) in days as the slope and the mid-transit time ($t_0$) as the y-intercept (similar to that done in \citealt{Turner2021,Turner2022}):

\begin{equation} \label{eq:phot_T(n)}
T(n) = t_0 + nP + \delta t_n,
\end{equation}
where $\delta t_n$ is the deviation of each midpoint from its expected value, otherwise known as the transit perturbation. This is commonly referred to as the observed minus calculated (O-C) value.

Because we solved for the impact parameter $b$ via the method detailed in \citet{Espinoza_2018}, the last parameter needed to calculate $a/R*$, the ratio of the semi-major axis and the stellar radius, is the stellar density $\rho_*$ \citet{Espinoza_2019}:

\begin{equation} \label{eq:a/R*}
    \frac{a}{R_*} = \left[\frac{\rho_* G P^2}{3\pi} \right]^{1/3},
\end{equation}
where $P$ is the orbital period.\texttt{juliet} then solves for inclination $i$ directly via the following equation (see, e.g., \citet{Winn_2010}, rearranged by \citet{Espinoza_2019}:

\begin{equation} \label{eq:inclination}
    i = \arccos \left[ \frac{b}{a/R_*} \left(\frac{1 + e \sin \omega}{1 - e^2}\right) \right],
\end{equation}
where $b$ is the impact parameter, $e$ is the eccentricity, and $\omega$ is the argument of periastron.

The nested sampling photometry model utilized the \texttt{dynamic dynesty} \citep{dynesty} sampler to model 3000 live points, which is within an acceptable order of magnitude of the \texttt{juliet} documentation recommendation. The \texttt{dynamic dynesty} sampler adjusted the number of samples dynamically during computation to improve accuracy and efficiency.  The model included a GP with a \texttt{celerite} approximate Matern kernel \citep{celerite} to fit and remove stellar variability, and residuals are consistent with a normal distribution. See Table \ref{tab:photom_run_numbers} for the priors and posteriors for the joint GP-photometry model.

We assessed the reliability of the GP fit by searching for evidence of overfitting and red noise in the residuals. Evidence of overfitting would be indicated by the noise of the residuals to be below the photon limit \citep{Swain2025}. The photon limit for $\pi$ Men is $\sim$10 ppm (cadence of 120 minutes, mag$_{TESS}$ = 5.11) and the root mean square in our residuals (bottom panel in Figure \ref{fig:stacked_phot}) is $\sim$200 ppm. Therefore, we are not overfitting our data. We searched for red noise in the residual light curve using the residual permutation \citep{Southworth2008}, time-averaging \citep{Pont2006}, and wavelet methods \citep{Carter2009} as implemented in the \texttt{EXOMOP} transit modeling code (\citealt{Pearson2014,Turner2016b}). Likewise, we do not find any red noise in the residuals. Based off these two tests, our GP fit is reliable.

\begin{deluxetable*}{lcc}
\label{tab:photom_run_numbers}
\tablecaption{Prior boundary conditions and resulting parameters from TESS}
\startdata
\tablehead{\colhead{Parameter}&\colhead{Prior or Fixed Value }&\colhead{Result}} 
P (days)  & --  & 6.267823$\pm$1$\times10^{-6}$\tablenotemark{$*$}   \\
$t_0$ (BTJD) & --  & 1325.5042$\pm$0.0003\tablenotemark{$*$}     \\
$\rho_*  (\text{kg}/\text{m}^3)$ & $\mathcal{J}[100.0,10000.0]$ & $1050^{+220}_{-230}$ \\
$r_1$ & $\mathcal{U}[0,1]$ & 0.74$\pm$0.05      \\
$r_2$ & $\mathcal{U}[0,1]$ & $0.0171^{+0.0003}_{-0.0002}$ \\
$R_p/R_*$ & --   & $0.0171^{+0.0003}_{-0.0002}$\tablenotemark{$\dagger$}  \\
$b$  & --   & $0.61^{+0.07}_{-0.08}$ \tablenotemark{$\dagger$}  \\
$a/R_*$  & --  & $13.0^{+0.9}_{-1.0}$\tablenotemark{$\diamond$}     \\
$i$ (\textdegree) & --  & $87.3^{+0.5}_{-0.6}$\tablenotemark{$\star$}   \\
$e$ & 0.0 & -- \\
$\omega$ & 90.0 & -- \\
$q_1$  & $\mathcal{N}[0.305, 0.1]$ & $0.29\pm0.04$ \\
$q_2$  & $\mathcal{N}[0.269, 0.1]$ & $0.27\pm0.08$  \\ 
$\sigma$  & $\mathcal{J}[1\times10^{-6}, 10^{6}]$  & $6.3\times10^{-5} \pm 0.1\times10^{-5}$ \\
$\rho$    &  $\mathcal{J}[0.001,1000.0]$ &  $0.236\pm0.007$ 
\enddata
\tablenotetext{$*$}{Directly computed via least-squares best fit to transit midpoint function $T(n)$, in which $P$ is the slope and $t_0$ is the intercept. See Equation \ref{eq:phot_T(n)}.}
\tablenotetext{\dagger}{Calculated from the posteriors of $r_1$ and $r_2$ \citep{Espinoza_2018}.}
\tablenotetext{\diamond}{Calculated from the posteriors of $\rho_*$ and $P$. See Equation \ref{eq:a/R*}.}
\tablenotetext{\star}{Calculated directly from impact parameter posterior. See Equation \ref{eq:inclination}.}
\end{deluxetable*}


\subsection{Photometry Results} \label{subsec:TTV_Results}

\begin{figure*}
  \begin{center}
    \includegraphics[width=\textwidth]{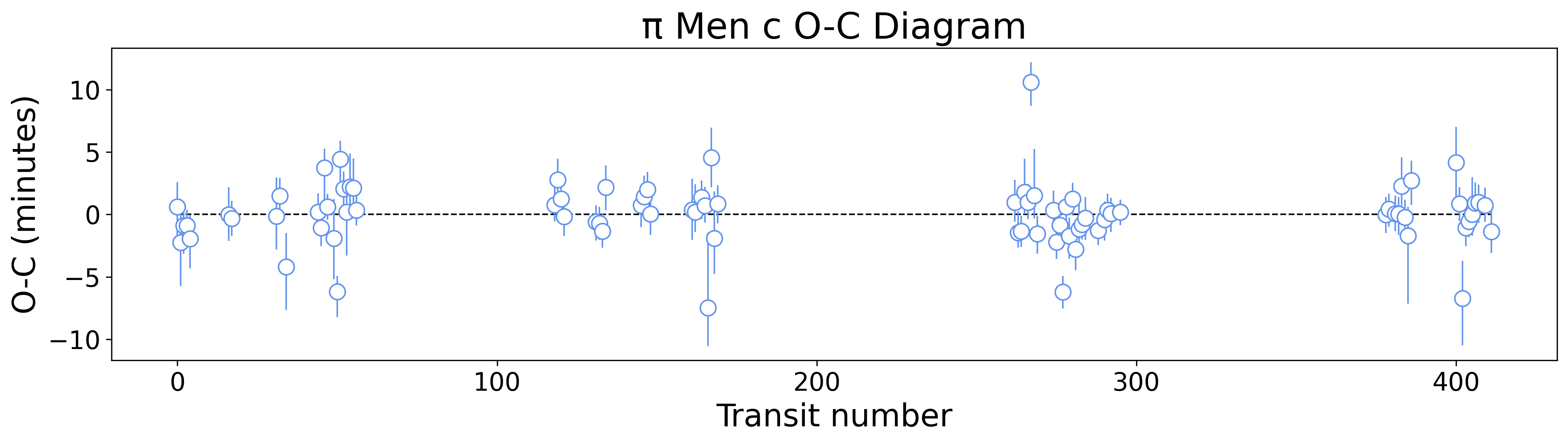} \\
    \caption{O-C Plot of $\pi$~Men~c transit midpoints. Midpoints are designated by epoch, and differences are in units of minutes. The dotted black line traces the path of expected transit midpoints values (in other words, the line where $\delta t_n$ = 0). We find our O-C plot to be consistent with a lack of TTVs.
    }
    \label{fig:oc}
  \end{center}
\end{figure*}

\begin{figure}
  \begin{center}
    \includegraphics[width=0.48\textwidth]{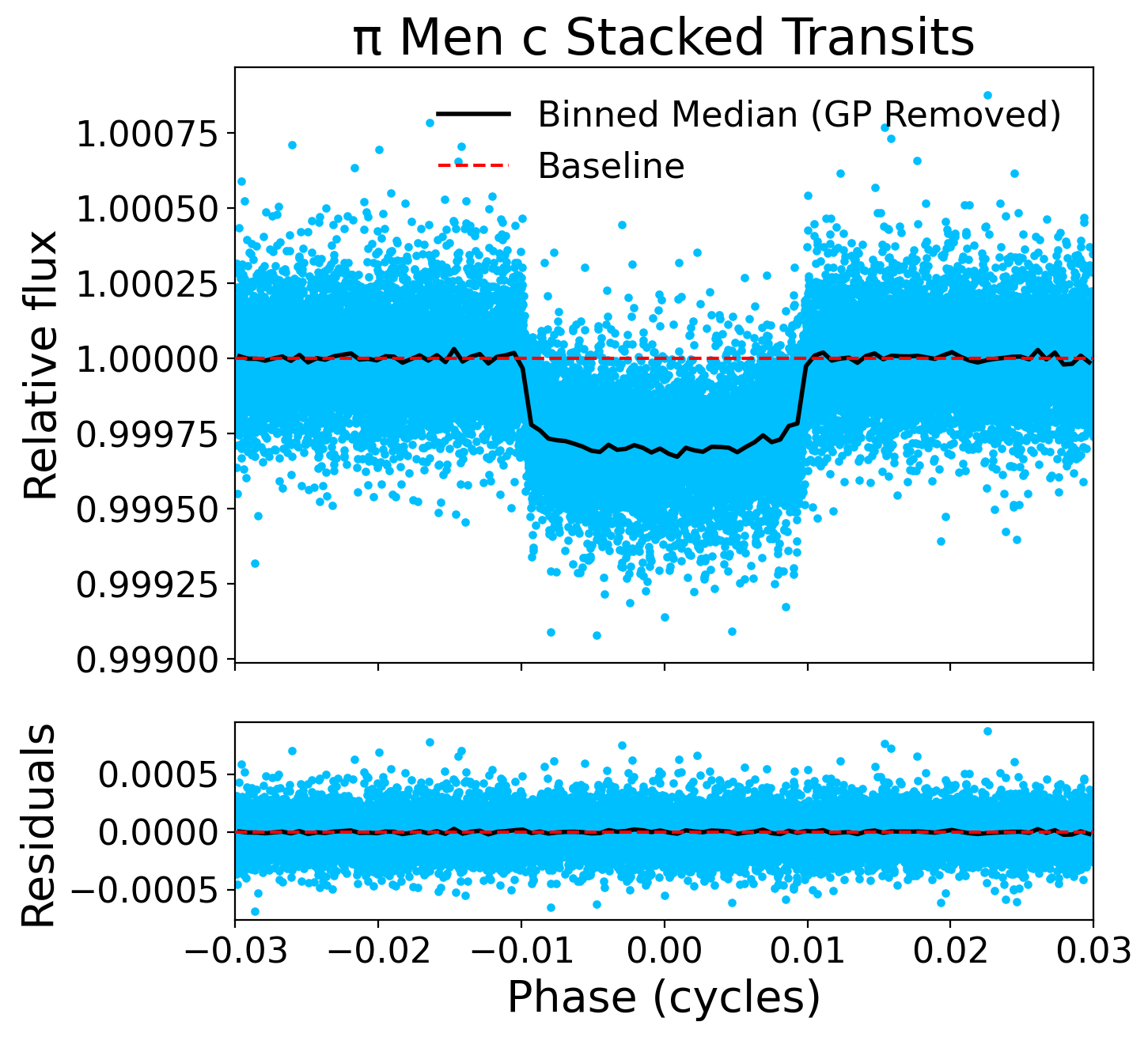} \\
    \caption{Phased-folded transit light curve of $\pi$~Men~c, where each transit is centered around their posterior transit midpoint. The GP Model has been subtracted from the data. The binned data ($n_{bins} = 100$) are shown as a black line. The residuals (light-curve minus transit+GP Model) are shown below the light curve, and are consistent with white noise. 
    }
    \label{fig:stacked_phot}
  \end{center}
\end{figure}

Table \ref{tab:photom_run_numbers} lists the planetary parameter priors and posteriors of the $\pi$~Men~c TTV photometry model, while Table \ref{tab:all_times} lists the posteriors of the transit midpoints. All derived $\pi$~Men~c planetary parameters are consistent with a Gaussian distribution; see the Appendix for associated corner plot. All the derived TESS transit times for  $\pi$~Men~c can be found in the Appendix (Table \ref{tab:all_times}). The observed minus calculated (O-C) transit times of $\pi$~Men~c TESS transits are shown in Figure \ref{fig:oc}. Figure \ref{fig:stacked_phot} shows the photometry phase curve of $\pi$~Men~c, with each stacked transit centered around its posterior midpoint.

We performed a reduced chi-squared test to calculate whether TTVs were present in the $\pi$~Men~c photometry scatter:

\begin{equation}
    \chi^2 = \frac{1}{N} \sum^N_{n=1} (\frac{\delta t_{n}}{\sigma_n})^2,
\end{equation}
in which N is the number of transits, $\delta t_{n}$ is the observed transit time minus the expected transit time of each transit n, and $\sigma_n$ is the $1\sigma$ error of each transit. Because each $\delta t_{n}$ has a high and low value, $\sigma_n$ is calculated by averaging the two errors for each point:

\begin{equation}
    \sigma_n = \frac{\sigma_{n, high} + \sigma_{n, low}}{2}.
\end{equation}

We adopt the TTV chi-squared threshold criteria utilized by \citet{Naponiello_2026}, which deems $\chi^2 \leq 2.0$ as timing scatter consistent with measurement errors. 

We use the \texttt{astropy.timeseries.LombScargle} class \citep{Astropy2022} to produce a frequency Lomb-Scargle periodogram and assess the periodicity of the O-C plot. Our $\Delta x$, the difference between each data point in the domain, is equal to 1 transit (or 1 epoch), and the total baseline is the transit number of the final transit minus the transit number of the first transit (411 transits). The minimum frequency is 1 divided by the transit number of the final transit, $f_{min} = 1/411 \sim 0.00243$, while the maximum frequency, otherwise known as the Nyquist frequency, is equal to $f_{max} = \frac{1}{2\Delta x} = 0.5$.

\begin{figure}
  \begin{center}
    \includegraphics[width=0.49\textwidth]{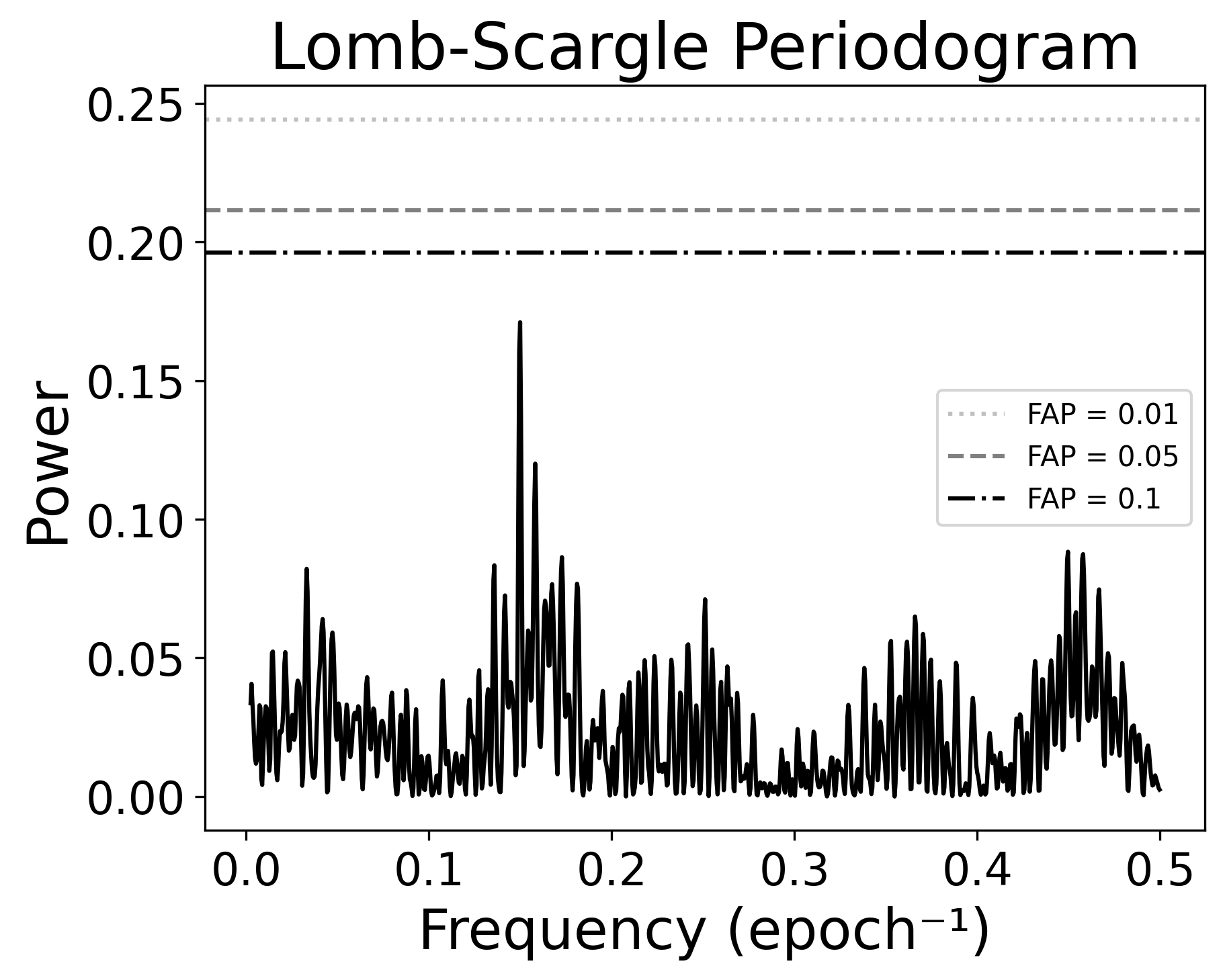} \\
    \caption{A frequency Lomb-Scargle periodogram of the $\pi$~Men~c O-C plot. The False-Alarm Probabilities (FAP) = 0.1, 0.05, and 0.01, measurements conditioned on the assumption of the null hypothesis of no signal, are shown as horizontal lines. Note that none of the periodogram peaks rise above any of the FAP thresholds, and as such, our O-C plot is consistent with a lack of TTVs.
    }
    \label{fig:lombscargle}
  \end{center}
\end{figure}

The observed minus calculated $\pi$~Men~c transit midpoints, shown in Figure \ref{fig:oc}, is consistent with the absence of TTVs ($\chi^2 \sim 1.64 \leq 2)$ in agreement with previous research \citep{Naponiello_2026}. In addition, no evidence of periodicity was found to be of greater significance than a False-Alarm Probability (FAP) = 0.1, our most generous significance threshold for a detectable signal. The highest Lomb-Scargle peak (Figure \ref{fig:lombscargle}), which occurred at a frequency of $f_{peak} \sim 0.15$ corresponding to a periodicity of $1/0.15 \sim 6.7$ transits, possessed a power of $\sim0.17$. Although lower than any FAP threshold, $f_{peak}$ reached a power value higher than all other local peaks. This is likely a product of the uneven transit midpoint sampling distribution. As such, both the chi-squared test and the Lomb-Scargle periodogram returned results consistent with a lack of TTVs.

\section{RADIAL VELOCITY}  \label{sec:RV}

\subsection{Radial Velocity Observations} \label{subsec:RV_Observe}

Our radial velocity measurements consist of archival data from many different publicly-available sources for maximum coverage. We used the High Accuracy Radial Velocity Planet Searcher (HARPS) HARPS-POST radial-velocity dataset from \citet{Hatzes_2022}; University College London Echelle Spectrograph (UCLES) measurements obtained by \citet{Jones_2002} and \citet{Butler_2006} as compiled by \citet{Gandolfi_2018}; and data from the Echelle SPectrograph for Rocky Exoplanets and Stable Spectroscopic Observations (ESPRESSO) and the CORALIE echelle spectrograph presented in \citet{Damasso_2020}. As in \citet{Damasso_2020}, the CORALIE dataset is divided into three different datasets, referred to as CORALIE-98, CORALIE-07, and CORALIE-14, to account for small RV offsets caused by significant instrument upgrades in 2007 and 2014. See the Appendix (Table \ref{tab:RV_datasets}) for a summary of RV datasets used; $N_{meas}$ is the number of RV measurements taken by each instrument. The data spans from January 1998 (UCLES) to December 2020 (HARPS). The complete RV data set sorted by source is given in the Appendix (Table \ref{tab:all_RV}).

Both the breadth and depth of the available radial velocity data are crucial for fully modeling the $\pi$ Men system. Though $\pi$~Men~c is the only transiting planet in the system, any radial velocity observations of $\pi$ Men will inevitably be dominated by the massive, highly eccentric, and non-transiting $\pi$~Men~b ($M_p = 12.6 M_{J}$, $P = 2088.33$ ± $0.34$ days, $e = 0.64$; \citealt{Hatzes_2022}). We provide models and updated parameters for both planets. The precise and high-density RV data of HARPS and ESPRESSO will constrain $\pi$~Men~c, but neither dataset fully encompasses one orbit of $\pi$~Men~b and thus cannot constrain $\pi$~Men~b's period. The UCLES and CORALIE data, though more sparse and less precise, span multiple decades and are crucial for constraining $\pi$~Men~b's entire orbit.

\subsection{Radial Velocity Data Analysis} \label{subsec:RV_DataAnalyis}

After using TESS photometry data to derive new orbital parameters for $\pi$~Men~c, the only transiting planet in the system, we then used those posteriors as priors in a new radial velocity nested sampling fit which obtained new parameters for each $\pi$ Men planet. We again used \texttt{juliet} for our fit, which in turn uses \texttt{RadVel} \citep{radvel} to model radial velocity time series. 

Multiple recent publications \citep{Hatzes_2022, Laliotis_2023, Harada_2025} have tentatively detected a third planet in the system, $\pi$~Men~d (P $\sim$ 120 days, e $>$ 0), via radial velocity measurements. To investigate this finding, we fit our RV data to both a 2-planet and a 3-planet model, the former modeling only $\pi$~Men~b and c, the latter model adding $\pi$~Men~d. The publicly-available radial velocity data described in Section \ref{subsec:RV_Observe} and Appendix Table \ref{tab:RV_datasets} were used to derive the following planetary parameters of $\pi$~Men~b, c, and d: radial velocity (K) in meters per second, period (P) in days, time of conjunction ($t_0$) in Barycentric Julian Date (BJD), eccentricity (e), and argument of periastron ($\omega$). See Table \ref{tab:RV_run_numbers} for the RV 2-planet and 3-planet model priors and posteriors. 

We again used \texttt{dynamic dynesty} as our sampler; the number of live points was increased to 20000 for the RV models to ensure clean posterior distributions. In both RV models, most priors were kept wide and uniform for maximum exploration of parameter space, other than the following exceptions. The $\pi$~Men~c period and $t_0$ were informed by the previously-derived photometry model; their priors are the derived values from the photometry model $P_c$ and $t_{0, c}$. Because $\pi$~Men~d is a relatively weak signal with few orbits of high-cadence data, our $t_{0, d}$ prior is a $5\sigma$ normal distribution around the $t_{0, d}$ posterior value reported by \citet{Harada_2025} instead of a uniform distribution.

\begin{deluxetable*}{lccc}
\tablewidth{\textwidth}
\label{tab:RV_run_numbers}
\startdata
\hline
\tablehead{\colhead{Parameter}&\colhead{Prior / Fixed Value }&\colhead{2-Planet Result}&\colhead{3-Planet Result}} 
$K_b$ (m/s) & $\mathcal{U}$[180.0,210.0] & 191.1$\pm$0.4 & 192.4$\pm$0.4 \\
$P_b$ (days)  & $\mathcal{U}$[2000,2300]  & 2088.8$\pm$ 0.4    &  2088.7$\pm$0.4\\
$t_{0, b}$ (BJD) & $\mathcal{U}${[2456528.0,2456568.0]}  & 2456534.5$\pm$0.8 & 2456537.9$\pm$ 0.8\\
$e_b$ & $\mathcal{U}$[0.3,0.8] & 0.6427$\pm$ 0.0007 &  0.6400$\pm$0.0007\\
$\omega_b$ (\textdegree) & $\mathcal{U}$[0.0,360.0] & 331.9$\pm$ 0.3 & 331.5$\pm$0.2\\
\hline
$K_c$ (m/s) & $\mathcal{U}$[0.0,100.0] & 1.39$\pm$0.09 & 1.16$\pm$0.08 \\
$P_c$ (days)  & $\mathcal{N}$[6.267823, 1$\times10^{-6}$]  & 6.267823$\pm$ 1$\times10^{-6}$    & 6.267823$\pm$ 1$\times10^{-6}$\\
$t_{0, c}$ (BJD) & $\mathcal{N}${[2458325.5042, 0.0003]}  & 2458325.5042$\pm$0.0003     & 2458325.5041$\pm$0.0003 \\
$e_c$ & $\mathcal{U}$[0.0,0.8] & 0.06$\pm$0.06 & $0.03^{+0.04}_{-0.02}$\\
$\omega_c$ (\textdegree) & $\mathcal{U}$[180.0,360.0] & 274$\pm$ 30 & $270^{+52}_{-48}$\\
\hline
$K_d$\tablenotemark{$*$} (m/s) & $\mathcal{U}$[0.0,100.0] & -- & 1.8$\pm$0.1\\
$P_d$\tablenotemark{$*$} (days)  & $\mathcal{U}$[110.0,130.0]  & --& 123.1$\pm$0.2\\
$t_{0, d}$\tablenotemark{$*$} (BJD) & $\mathcal{N}${[2455218.6,20.0]}  & --    & 2455259$\pm$6 \\
$e_d$\tablenotemark{$*$} & $\mathcal{U}$[0.0,0.8] & --& 0.41$\pm$0.05\\
$\omega_d$\tablenotemark{$*$} (\textdegree) & $\mathcal{U}$[180.0,360.0] & --& $353^{+5}_{-6}$
\tablecaption{Prior boundary conditions and resulting parameters from Radial Velocity}
\enddata
\tablenotetext{$*$}{$\pi$~Men~d parameters were only calculated in the 3-planet model.}
\end{deluxetable*}

\subsection{Radial Velocity Results} \label{subsec:RV_Results}

Table \ref{tab:RV_run_numbers} lists the priors and posteriors of both the 2-planet and 3-planet radial velocity models. Both models had the same priors, with the only exception being the addition of the third planet $\pi$~Men~d parameters in the 3-planet model.

To assess the quality of both the 2-planet and 3-planet radial velocity models, we assess the log evidences, $\ln \mathcal{Z} = \ln \mathcal{P}(D|$Model), of each model with respect to the data $D$. If both models are equally likely (i.e. $\mathcal{P}(\text{Model}_i)/\mathcal{P}(\text{Model}_j) = 1$), the odds ratios derived from the log-evidence of Model$_i$ and Model$_j$ are calculated by subtracting one log-evidence from the other:

\begin{equation}
    \ln \frac{\mathcal{P}(\text{Model}_i | D)}{\mathcal{P}(\text{Model}_j | D)} =
    \ln \frac{\mathcal{P}(D | \text{Model}_i)}{\mathcal{P}(D | \text{Model}_j)} = 
    \ln \frac{\mathcal{Z}_i}{\mathcal{Z}_j},
\end{equation}

\begin{equation}
    |\ln \frac{\mathcal{Z}_i}{\mathcal{Z}_j}| = |\ln \mathcal{Z}_i - \ln \mathcal{Z}_j| = \Delta \ln \mathcal{Z}.
\end{equation}

The magnitude of $\Delta \ln \mathcal{Z}$ reflects how strongly the data prefer one model over the other—the larger $\Delta \ln \mathcal{Z}$ is, the more the model with the higher log evidence is favored relative to its competitor. We find a log-evidence from the 2-planet model ($\ln \mathcal{Z}_2$) of $-1725.259 \pm 0.131$ and a log-evidence from the 3-planet model ($\ln \mathcal{Z}_3$) of $-1635.374 \pm 0.072$. Thus, $\Delta \ln \mathcal{Z} = 89.605$ implies a odds ratio substantially in favor of the 3-planet model.

\begin{figure*}
  \begin{center}
    \includegraphics[width=\textwidth]{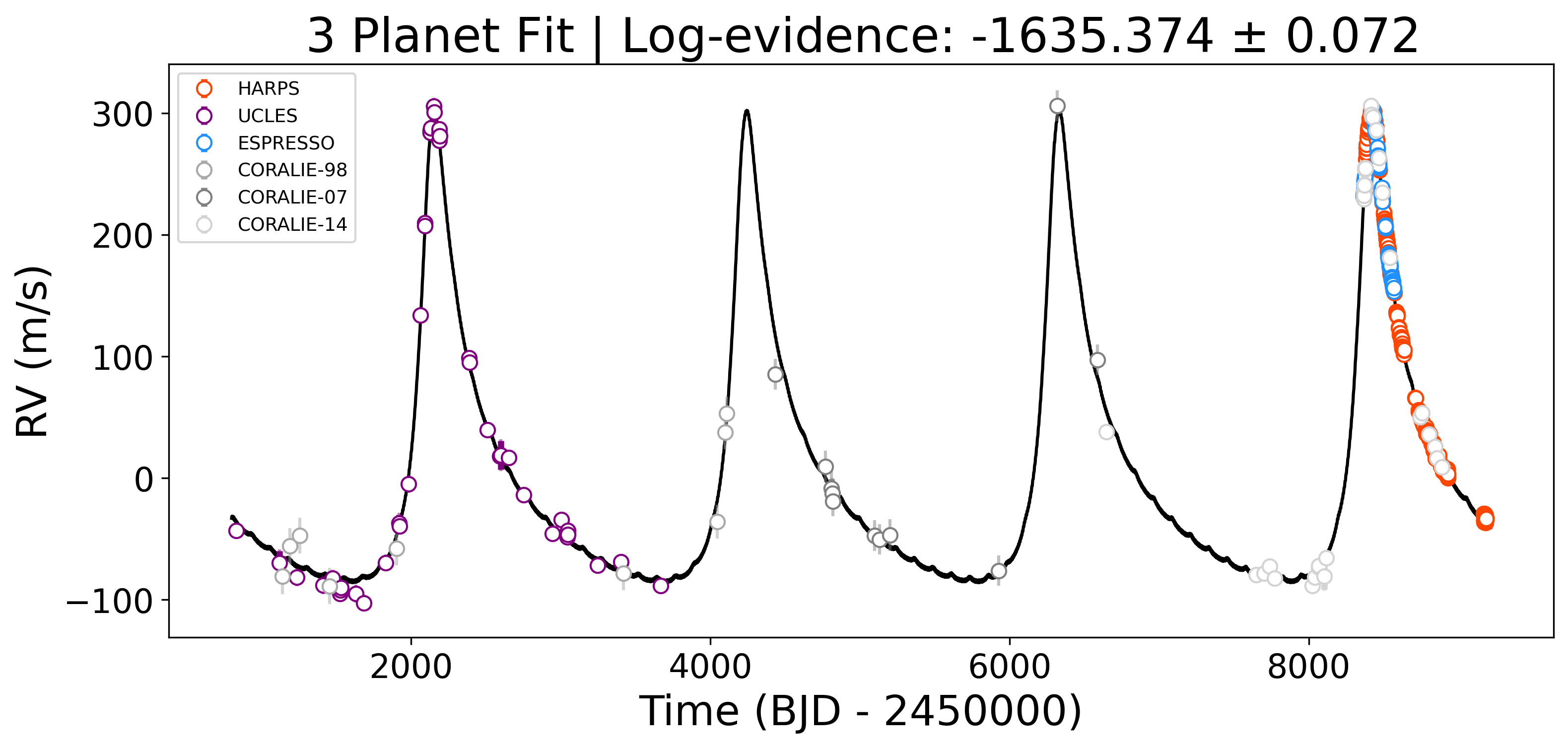} \\
    \caption{The 3-planet nested sampling model fit of the $\pi$ Men system radial velocity data. The log-evidence of the 3-planet model was more favored than the 2-planet model.
    }
    \label{fig:3planet_RV}
  \end{center}
\end{figure*}

\begin{figure}
  \begin{center}
    \includegraphics[width=0.45\textwidth]{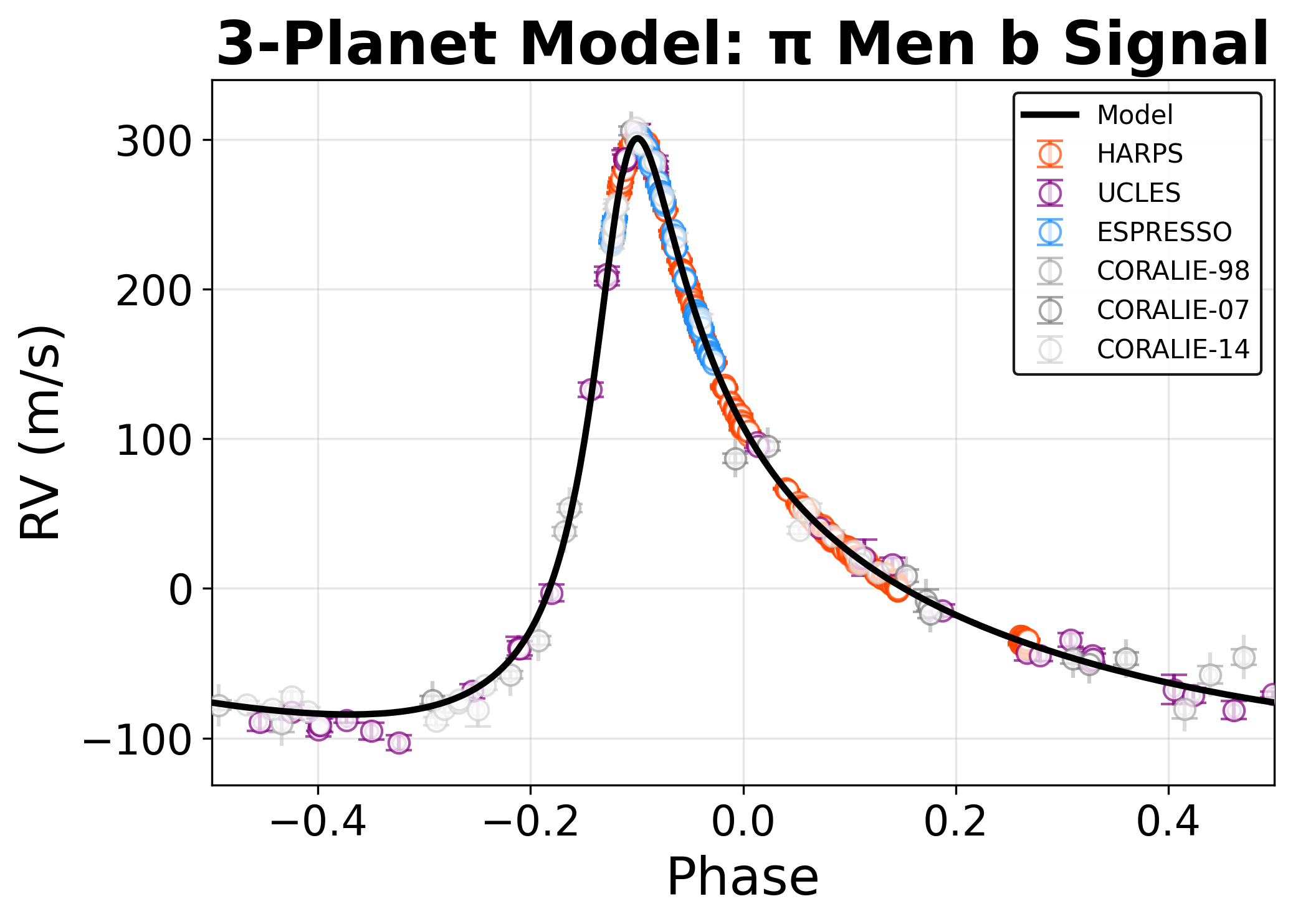} \\
    \includegraphics[width=0.45\textwidth]{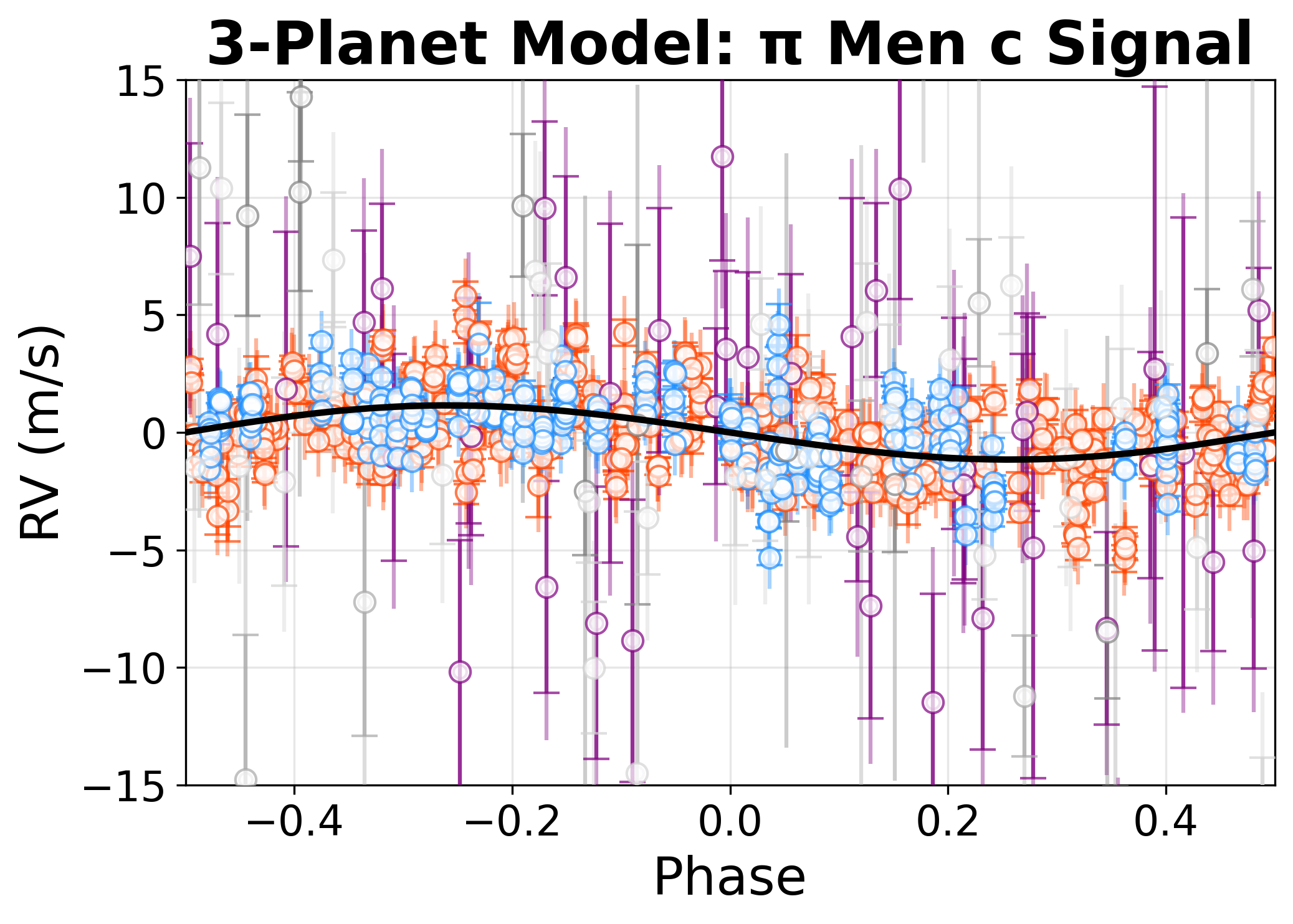} \\
    \includegraphics[width=0.45\textwidth]{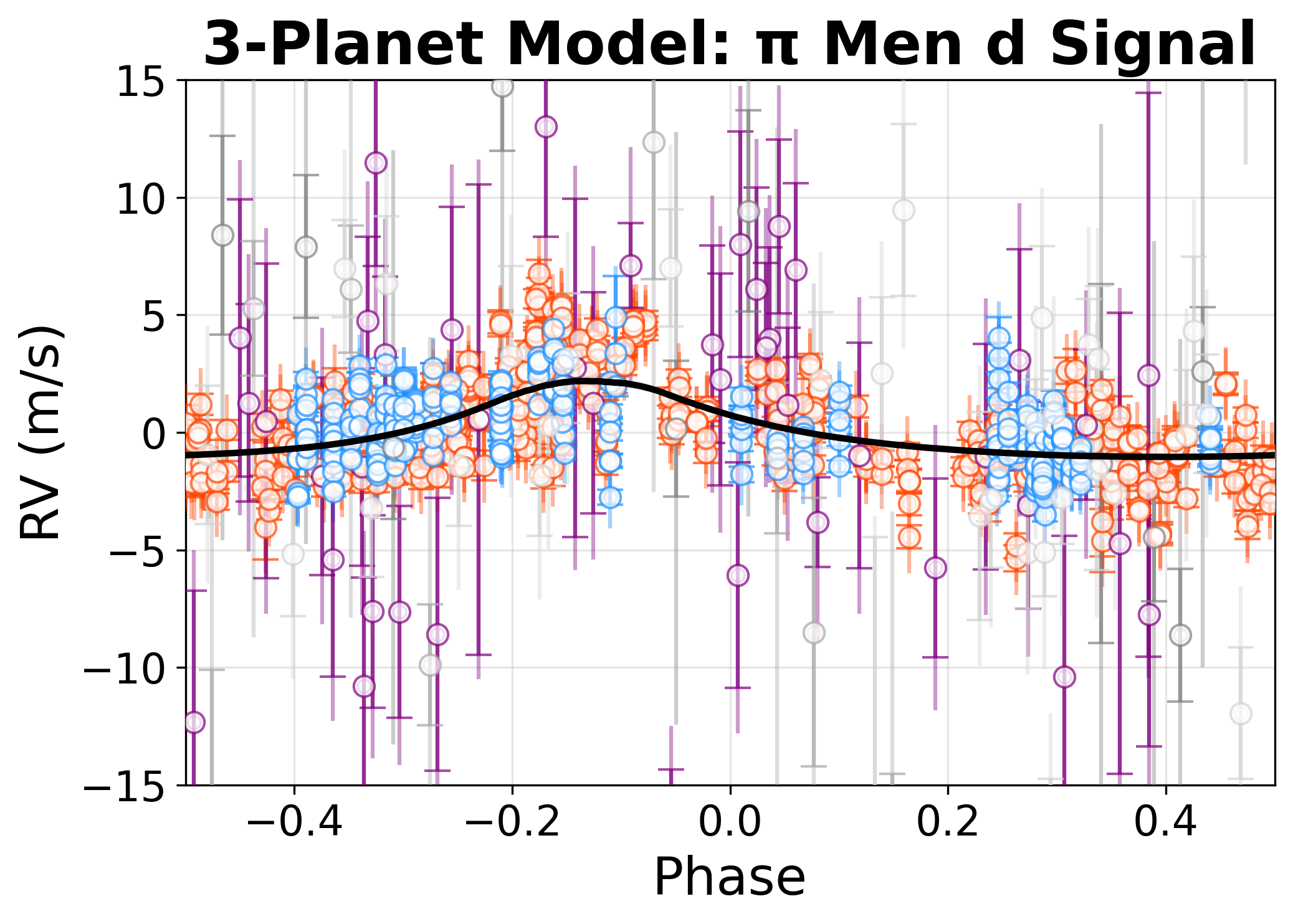} \\
    \caption{3-planet RV model phase curves. Top: Radial velocity phase curve of the massive and highly eccentric $\pi$~Men~b. Middle: Radial velocity phase curve of $\pi$~Men~c, a much smaller signal. Bottom: Radial velocity phase curve of $\pi$~Men~d, a signal of similar amplitude to $\pi$~Men~c, with a much wider period and higher eccentricity.}
    
    \label{fig:RV_phasecurves}
  \end{center}
\end{figure}

\begin{figure}
  \begin{center}
    \includegraphics[width=0.45\textwidth]{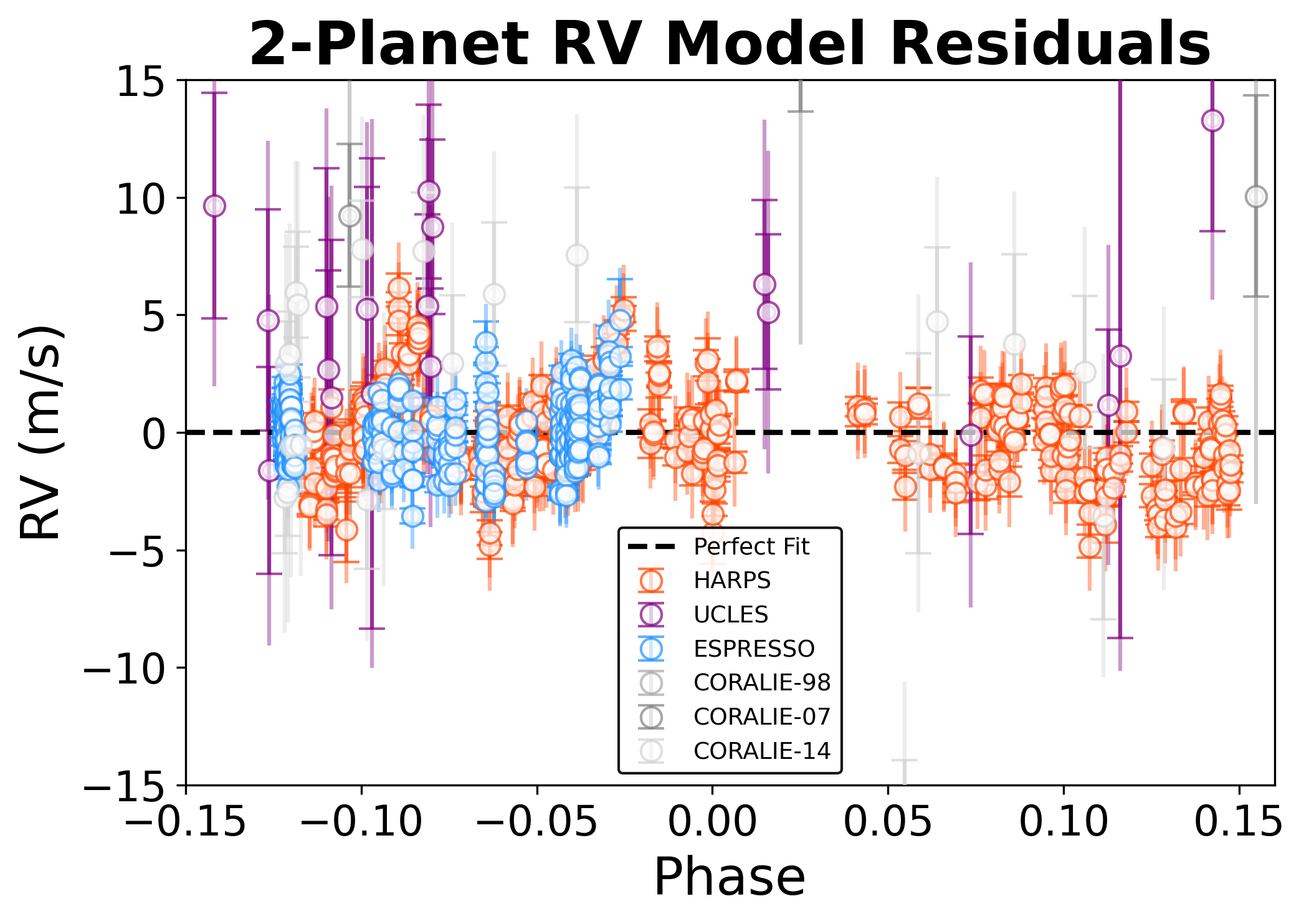} \\
    \includegraphics[width=0.45\textwidth]{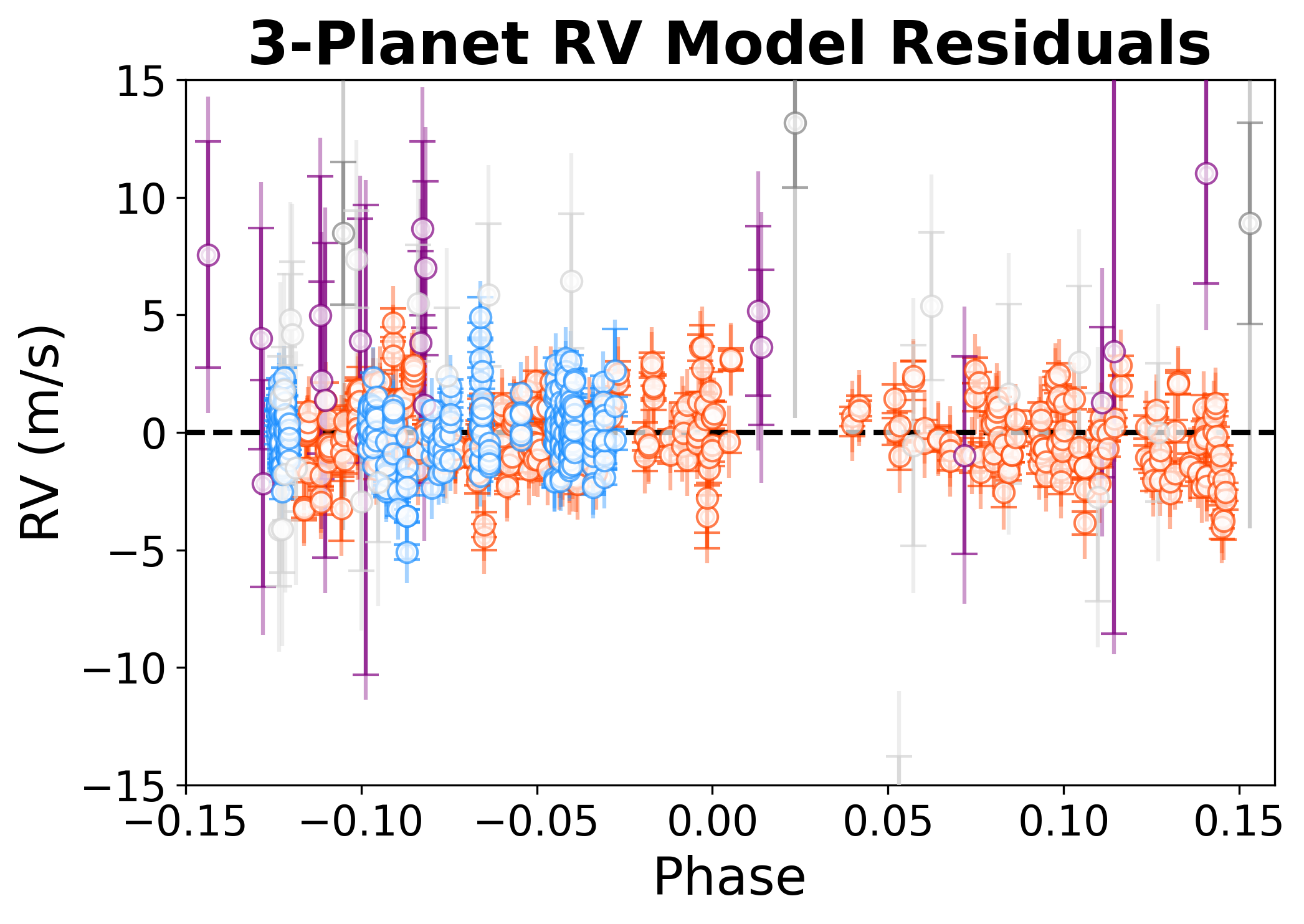} \\
    \caption{RV model residuals, zoomed in on the most concentrated portion of the HARPS and ESPRESSO time series. Top: The residuals of the 2-planet model. Bottom: The residuals of the 3-planet model. The residuals of the 3-planet model are more flat and statistically preferred than the 2-planet model.
    }
    \label{fig:RV_residuals}
  \end{center}
\end{figure}

See Figure \ref{fig:3planet_RV} for a complete depiction of the 3-planet model compared with the data, and Figure \ref{fig:RV_phasecurves} for the 3 phase curves of $\pi$~Men~b, c, and d according to the 3-planet model. Figure \ref{fig:RV_residuals} compares the residuals between both models. Note that in Figures \ref{fig:3planet_RV}, \ref{fig:RV_phasecurves}, and \ref{fig:RV_residuals}, the data points have two sets of error bars each: the error bars with brackets are the raw errors $\sigma(t)^2$ calculated by \texttt{juliet}, while the error bars without brackets take an instrument-based jitter term $\sigma_{w, i}^2$ into account to create a noise model $\epsilon_i (t)$ for instrument $i$:

\begin{equation}
    \epsilon_i (t) \sim \mathcal{N}(0, \sqrt{\sigma(t)^2 + \sigma_{w, i}^2} ).
\end{equation}

The designation of a separate parameter $\epsilon_i(t)$ is meant to provide a quantifiable comparison between instrument precisions.

We calculate the mass of $\pi$~Men~c from our radial velocity and photometry posterior estimates via the rearranged Equation 14 from \citet{Lovis_2010}, assuming $M_p << M_*$:

\begin{equation} \label{eq:planet_mass}
    M_p = \frac{1}{\sin{i}} \frac{KM_{J}\sqrt{1 - e^2}}{28.4325 \text{m/s}} (\frac{M_*}{M_\odot})^{2/3}(\frac{P}{1 \text{yr}})^{1/3},
\end{equation}
where $K$ is the radial velocity in m/s, $M_{J}$ is the mass of Jupiter in any desired units, $e$ is the eccentricity, $M_*$ and $M_{\odot}$ are the stellar and solar masses respectively, and $P$ is the planet period in years. We find that the mass of $\pi$~Men~c is $M_c = 0.0110\pm0.0008 M_{J} = 3.50\pm0.25 M_E$, which is within the uncertainties of previous estimates. Our result increases $M_c$ estimate precision by a factor of 1.4. This new mass estimate will help further constrain the properties of $\pi$~Men~c, including its escape velocity and visibility via transmission spectroscopy.

\section{DISCUSSION} \label{sec:discussion}

\subsection{$\pi$~Men~c Characterization} \label{subsec:disc_1}

\begin{figure*}
  \begin{center}
    \includegraphics[width=\textwidth]{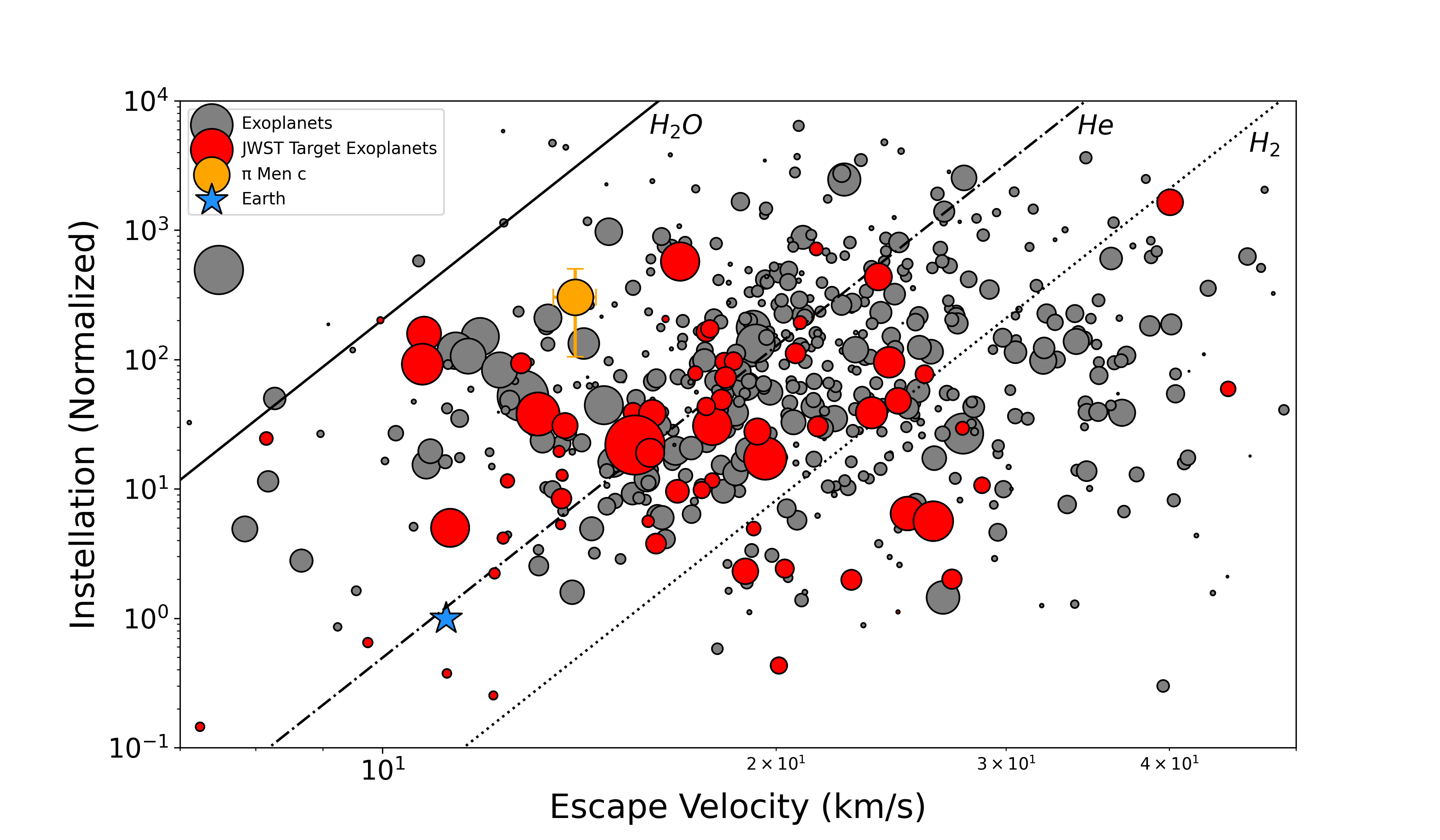} \\
    \caption{The logarithmic Instellation (normalized to Earth) vs. Escape Velocity of 564 cataloged exoplanets. JWST Targets as of Cycle 5 are highlighted in red, while our $\pi$~Men~c measurements and propagated uncertainties are orange. Earth's position on this relation is designated as a blue star. The size of each point corresponds to its Transmission Spectroscopy Meric (TSM), a value proportional to the expected signal-to-noise ratio of a transmission spectra. $\pi$~Men~c occupies a unique space within escaping atmospheres while possessing a TSM value comparable to JWST targets.}
    \label{fig:shoreline_complex}
  \end{center}
\end{figure*}

We used both a reduced chi-squared test and a Lomb-Scargle periodogram to assess the potential periodicity of $\pi$~Men~c transit midpoints. Both methods returned results consistent with an absence of TTVs. We used this finding to inform our further analysis of $\pi$~Men~c.

$\pi$~Men~c is unique as a sub-Neptune planet with an instellation and escape velocity ($v_{esc}$) that would suggest it may lose its lighter atmospheric gases. Figure \ref{fig:shoreline_complex} depicts the escape velocities vs. normalized instellations for 564 confirmed exoplanets, with the data taken from TEPCat\footnote{\cite{TEPCat_2011}, \url{https://www.astro.keele.ac.uk/jkt/tepcat/allplanets-noerr.html}}. The sizes of the graph points display the Transmission Spectra Metrics (TSMs) of each exoplanet, as described by \citet{Kempton_2018} and detailed later in this section. All accepted JWST Targets up to Cycle 5, sourced from TrExoLiSTS\footnote{\cite{Nikolov_2023}, \url{https://www.stsci.edu/~nnikolov/TrExoLiSTS/JWST/trexolists.html}}, are marked in red, and our $\pi$~Men~c estimate is shown in orange. 

Our estimate for $\pi$~Men~c's equilibrium temperature ($T_{eq}$) was calculated using a modified version of \citet{Southworth_2010} Equation 4 that assumes the bond albedo is equal to one minus the original equation's heat redistribution factor (e.g., Equation 7 from \citealt{Turner_2013}):
\begin{equation}
    T_{eq} = T_{eff} \left ( \frac{1}{4} \right )^{\frac{1}{4}} \left ( \frac{R_*}{2a_p} \right )^{\frac{1}{2}},
\end{equation}
where $T_{eff}$ is the effective temperature of the star as reported by \citet{Hatzes_2022}, and $a_p$ is the planetary semimajor axis. We find our newly derived equilibrium temperature to be $T_{eq} = 1176$ K, which is well within previous estimates.

Using our observed value of planetary mass and our derived value of equilibrium temperature, the escape velocity ($v_{esc}$), instellation ($I$), and Transmission Spectroscopy Metric (TSM) are calculated as follows \citep{Zahnle_Catling_2017, Kempton_2018}:

\begin{equation}
    v_{esc} = \sqrt{\frac{2 GM_{p}}{R_{p}}},
\end{equation}

\begin{equation}
    I = \frac{L_*}{L_\odot} \times (\frac{a_E}{a_p})^2,
\end{equation}

\begin{equation}
    TSM = S \times \frac{R_p^3 T_{eq}}{{M_p R_*^2}} \times 10^{\frac{-m_K}{5}},
\end{equation}
where $L_*$ is the host star luminosity, $L_\odot$ is the luminosity of the Sun, $a_E$ is the semimajor axis of the Earth, $a_p$ is the exoplanet semimajor axis, $m_K$ is the apparent magnitude of the host star in the K band, and $S$ is a scalar that depends on $R_p$ \citep{Kempton_2018}. The TSM is meant to be proportional to the expected signal-to-noise ratio from a transmission spectra based on planetary properties and assuming a cloud-free atmosphere \citep{Kempton_2018}; thus, planets with larger TSM values will be higher-priority targets for observations and modeling. Note that our TSM equation is slightly modified from \citet{Zahnle_Catling_2017}: we use K magnitudes, which we source from TEPCat\footnote{\cite{TEPCat_2011}, \url{https://www.astro.keele.ac.uk/jkt/tepcat/observables.html}},  instead of J magnitudes due to bulk data availability/comparability. Additionally, any exoplanet with R $>$ 10R$_{\oplus}$, thus having a non-physical TSM value according to Figure 5 of \cite{Kempton_2018}, is excluded from Figure \ref{fig:shoreline_complex}.

Our estimates of $\pi$~Men~c $v_{esc}$ and Normalized $I$, using the $M_*$ and $R_*$ values reported by \citet{Hatzes_2022} and $L_*$ reported by GAIA \citep{GAIA_2016, GAIADR3_2023}, are $v_{esc, c} = 14.0\pm0.5$ km/s and $I_c = 303^{+198}_{-199}$. Our values are consistent with previous publications and improve the precision of previous $v_{esc, c}$ and Normalized $I_c$ estimates, but predictably, the limiting factors for increasing precision in both values are the stellar parameters. Improving the uncertainties of the stellar parameters would greatly increase the certainty of $v_{esc}$ and $I$, and as such we recommend that future observations focus on observing stellar parameter/property estimates at higher precision. 

\citet{Kempton_2018} considers a significantly visible TSM for a small sub-Neptune ($1.5 < R_p < 2.75$) to be $TSM > 90$; any candidates above this threshold are deemed high-priority targets for ground-based and space-based transmission spectra observations. We calculate a value of $TSM_c \sim 464$ for $\pi$~Men~c using our version of the TSM equation (K magnitudes instead of J magnitudes). Substituting in the $\pi$~Men~c J magnitude instead returns $TSM_c \sim 347$; though a smaller value, it is still easily above the high-priority threshold set by \citet{Kempton_2018} and comparable to the TSM values of exoplanet JWST targets. 

Observing high-precision transmission spectroscopy of $\pi$~Men~c with JWST would answer many questions about not only its current conditions, but how it formed. Because $\pi$~Men~c is predicted to retain heavier molecules such as H\textsubscript{2}O and CO\textsubscript{2}, determining whether these species are present or absent would clarify whether $\pi$~Men~c formed beyond the snow line and subsequently migrated inward, and whether its atmospheric properties align with theoretical expectations and with those of other steam worlds \citep{García_Muñoz_2021, Ghorayeb_2024}. Even a flat transmission spectrum would be informative. Given its low density, $\pi$~Men~c cannot be an airless rock \citep{Huang_2018, Gandolfi_2018}. Clouds can mute molecular features and produce a flat spectrum, but they are unlikely on a $T_{eq}>700$ K sub-Neptune \citep{Brande_2024}, and our value of $\pi$~Men~c’s $T_{eq} = 1176$ K far exceeds this. \citet{García_Muñoz_2021} observes that C II ions are actively escaping $\pi$~Men~c's atmosphere, which would make a flat spectrum unlikely. But if the spectrum is found to be flat, \citet{García_Muñoz_2021} predicts a high–mean-molecular-weight atmosphere, a small scale height, and that $\pi$~Men~c will retain its current atmosphere and radius. In all cases, the transmission spectrum will refine our understanding of the formation and evolution of the $\pi$ Men system and similar systems.

Not only is $\pi$~Men~c highly visible via transmission spectroscopy, it occupies a unique atmospheric escape status compared to most other known exoplanets (Figure \ref{fig:shoreline_complex}). It exists squarely between the H\textsubscript{2}O and He molecular escape velocity zones in a relatively under-sampled and unexplored region with few other high-TSM exoplanets. In addition, its status as planet within a keystone radial velocity system represents a unique opportunity to characterize even more of its parameters and planetary properties. For all of the above reasons, we deem $\pi$~Men~c a high-priority target for transmission spectra measurements, particularly from the JWST.

\subsection{$\pi$ Men d, the Third Planet?}
\label{subsec:disc_2}

We utilized both 2-planet and 3-planet models to investigate the previously reported possibility of a third planet in the $\pi$ Men system, $\pi$~Men~d. The 3-planet model was found to be substantially preferred over the 2-planet model, suggesting that $\pi$~Men~d is statistically favored to exist. We find the signal associated with $\pi$~Men~d to be consistent with a moderately eccentric ($e_d = 0.41\pm0.05$) planet with a period of $P_d = 123.1\pm0.2$ days. The radial velocity amplitude ($K_d = 1.8\pm0.1$ m/s) is similar in size to the signal produced by $\pi$~Men~c, but its period, which is larger by multiple orders of magnitude, implies a higher mass. 

By combining our posterior parameter results and the boundary conditions inferred by previous research, we can derive a broad mass range for $\pi$~Men~d. \citet{Hatzes_2022} found that the mass of $\pi$~Men~d would need to be $<$ 20 M$_{\oplus}$ to maintain a stable orbit, and their orbital solution implies the inclination to be $i_d >$ 40\textdegree. Because $\pi$~Men~d is non-transiting, we can further assume that $i_d <$ 85\textdegree. Given all of these boundary conditions, we find using Equation \ref{eq:planet_mass} and our estimates of the planetary parameters that the possible mass range of $\pi$~Men~d is 13.4 $\leq M_d <$ 20 M$_{\oplus}$. This places $\pi$~Men~d in the sub-Neptune to Neptune-like mass range, if the planet truly exists.

We note that we were able to constrain $\pi$~Men~d's time of conjunction ($t_{0,d}$) by fitting it to a normal prior on this parameter instead of a uniform prior. If the $t_{0,d}$ fit was allowed to be uniform, the posterior would approach the upper boundary condition, regardless of how wide the boundary conditions were. We attribute this to the relatively weak signal of $\pi$~Men~d. The other two planets in the system are far more easily characterized: $\pi$~Men~b's signal is unmistakably large and most of its orbit is characterized by at least one instrument, and though $\pi$~Men~c's amplitude is small, it has both a wealth of photometry data and many high-cadence RV periods to its name. Our knowledge and characterization of $\pi$~Men~d, a small-amplitude and non-transiting planet with only 2-3 total periods of RV data, would greatly benefit from more full periods of high-cadence data on par with the precision of the already-existing HARPS and ESPRESSO $\pi$ Men observations.

\subsection{Dynamics of the $\pi$ Men System}
\label{subsec:disc_3}

The absence of detectable TTVs for $\pi$~Men~c gives an important dynamical verification of the radial-velocity architecture investigated in this work. The observed O-C scatter is consistent with measurement uncertainties, with a reduced $\chi^2 \simeq 1.64$, and the Lomb-Scargle periodogram shows no clear signal. This is expected for the best-fit three-planet architecture because neither $\pi$~Men~b nor the candidate $\pi$~Men~d lies near a low-order mean-motion resonance with $\pi$~Men~c.

We present the following calculation to assess whether $\pi$~Men~d can induce TTVs in the $\pi$ Men system. To start, the period ratio between $\pi$~Men~d and $\pi$~Men~c is ${P_d}/{P_c} \simeq 19.6,$ far from resonant configurations that would typically generate large TTV signals (e.g. \citealt{Agol2005,Steffen2012,Dawson2019,ELMoutamid2023}). A simple estimate of the short-period perturbation \citep{Agol2005} caused by $\pi$~Men~d on $\pi$~Men~c would be:
\begin{equation}
\delta t_c \sim \frac{P_c}{\pi} \frac{M_d}{M_*}\left(\frac{a_c}{a_d}\right)^3.  
\end{equation}
Using the parameters from the photometry transit model and the three-planet radial-velocity solution (see Tables \ref{tab:photom_run_numbers} and \ref{tab:RV_run_numbers}) and the stellar parameters from Table \ref{tab:planet_parameters}, this gives $\delta t_c \sim 0.02$ s for the nominal solution. This is orders of magnitude below the timing precision of the TESS transit midpoints ($\sim$25 seconds), and therefore the lack of detectable TTVs is dynamically consistent with the proposed $\pi$~Men~d signal. 


The candidate third planet is also well separated from $\pi$~Men~c in terms of the mutual Hill radii. The mutual Hill radius \citep{Hill1878,Chambers1996} is defined as: 
\begin{equation}
R_{\rm H,mut} =  \left(\frac{M_c+M_d}{3M_*}\right)^{1/3} \frac{a_c+a_d}{2}.
\end{equation}
Using the planetary parameters in Tables \ref{tab:photom_run_numbers} and \ref{tab:RV_run_numbers}, this gives $R_{\rm H,mut}\simeq 0.0071~{\rm au}$. The orbital separation is therefore $\Delta_{c,d} = ({a_d-a_c})/{R_{\rm H,mut}} \simeq 60,$ which is much larger than the classical two-planet Hill-stability limit \citep{Chambers1996} of $2\sqrt{3}$. Even accounting for the eccentricity of $\pi$~Men~d, its pericenter distance is $d_p~=~a_d(1-e_d) \simeq 0.29~{\rm au},$ well exterior to the orbit of $\pi$~Men~c. Thus, the inner pair is expected to be strongly Hill stable.


The dynamical relationship between $\pi$~Men~d and the massive eccentric outer companion $\pi$~Men~b is potentially more important. For $P_b \simeq 2089$ days and $e_b \simeq 0.64$, the semimajor axis and pericenter distance of $\pi$~Men~b are approximately $a_b~\simeq~3.3~{\rm au}$, $\qquad~d_p ~\simeq~1.19~{\rm au}$. The apocenter distance of $\pi$~Men~d is $a_p = a_d(1+e_d) \simeq 0.70~{\rm au},$ leaving a radial gap of about $0.5$ au between $\pi$~Men~d at apocenter and $\pi$~Men~b at pericenter. The separation between $\pi$~Men~d and $\pi$~Men~b is therefore not orbit crossing for the nominal solution, although the large mass and eccentricity of $\pi$~Men~b imply that secular perturbations could still affect the long-term eccentricity and apsidal evolution of $\pi$~Men~d.

These simple estimates indicate that the three-planet solution is dynamically plausible at the level of orbital spacing and expected TTV amplitudes. However, because $\pi$~Men~b is massive and eccentric and because the estimated/measured eccentricity of $\pi$~Men~d is small, a full long-term $N$-body stability analysis would be valuable. Such a detailed $N$-body simulation is beyond the scope of this paper, but such an endeavor would prove fruitful in further research. Such integrations should explore the posterior distributions of $e_d$, $\omega_d$, and $M_d \sin i_d$, as well as possible mutual inclinations, in order to determine whether the statistically favored three-planet solution remains stable over secular and Gyr timescales.







\section{CONCLUSION} \label{sec:conclusion}

In this study, we applied photometric and radial velocity models to 20+ years of observations and inferred the orbital and planetary parameters for up to three planets in the $\pi$ Men system. Our analysis combines 6 years of publicly available TESS photometry of the transiting planet $\pi$~Men~c with 22 years of radial velocity observations of $\pi$ Men, and the photometry model was used to inform the RV fit. 

We calculated an O-C plot of $\pi$~Men~c's transit midpoints, and determined via reduced chi-squared test and Lomb Scargle periodogram that our data were consistent with a lack of transit timing variations. Our polished planetary parameters for the $\pi$~Men system include a period estimate for $\pi$~Men~c that is an order of magnitude more precise than in previous research.  $\pi$~Men~c exists in an under-explored region of atmospheric escape as a near-radius gap gaseous planet thought to lose its primary atmosphere but keep heavier elements, and our derived parameters including $v_{esc}$, $I$ and TSM suggest that $\pi$~Men~c would be a highly-informative target for transmission spectroscopy studies. We advise that further observations prioritize constraining the parameters of the host star $\pi$ Men, and we advocate for future transmission spectroscopy studies, such as via the JWST.

We refined the planetary and orbital parameters for every known planet in the $\pi$ Men system, and explored the possibility suggested by previous research of a third planet $\pi$~Men~d. Our radial velocity model log-evidences highly favor a 3-planet system and our results are consistent with the existence of $\pi$~Men~d (13.4 $\leq$ M$_d$ $<$ 20 M$_{\oplus}$). More high-cadence radial velocity data are required to better confirm the existence of and constrain the planetary properties, particularly the time of conjunction, of this third planet. The $\pi$ Men system is meaningful for the advancement of orbital dynamics and exoplanet atmospheres, and further study via ground-based and space-based observations will be very fruitful.

\begin{acknowledgments}
We would like to thank N\'{e}stor Espinoza and Leslie Hebb for their valuable insights, which greatly improved the paper.

J.D.T. was supported by NASA through Grant/Contract No. G06165 issued through the TESS General Investigator Program. L.A. is supported by the Cornell College of Arts \& Sciences Klarman Postdoctoral Fellowship. 
D.A.Y. is supported by a Juan Carlos Torres Postdoctoral Fellowship at the Massachusetts Institute of Technology. D.S-O. is supported by the National Science Foundation Graduate Fellowship Program, Grant No. DGE-2139899. 

This research has made use of the NASA's Astrophysics Data System Bibliographic Services and the the NASA Exoplanet Archive, which is operated by the California Institute of Technology, under contract with the National Aeronautics and Space Administration under the Exoplanet Exploration Program.  

This paper includes data collected by the TESS mission, which are publicly available from the Mikulski Archive for Space Telescopes (MAST). Funding for the TESS mission is provided by NASA's Science Mission Directorate. All the TESS data used in this paper can be found in MAST.  

We thank the anonymous referee for their helpful comments. 

\end{acknowledgments}

\facilities{TESS \citep{Ricker2015}; Exoplanet Archive \citep{Christiansen2025, ps}; UCLES \citep{Diego_1990}; HARPS \citep{HARPS_2003}; ESPRESSO \citep{ESPRESSO_2021}; CORALIE \citep{CORALIE_2000}}

\software{\texttt{astropy} \citep{astropy2013,astropy2018,Astropy2022}; \texttt{scipy} \citep{scipy_2020};
\texttt{juliet} \citep{Espinoza_2019}; \texttt{lightkurve} \citep{lightkurve_2018}; \texttt{batman} \citep{batman}; \texttt{ExoTiC-LD} \citep{Grant_2024}; \texttt{EXOMOP} \citep{Pearson_2014, Turner_2016}; \texttt{scipy} \citep{scipy_2020}; \texttt{celerite} \citep{celerite}, \texttt{dynamic dynesty} \citep{dynesty}; \texttt{RadVel} \citep{radvel}}

\appendix

Table \ref{tab:Phot_sectors} lists the sectors of TESS data which contained $\pi$~Men~c transits, along with the number of transits per sector and the sectors' start and end dates. The only sector excluded from this table was Sector 27 (s027): $\pi$ Men was technically within TESS's field of view during this sector, but all transits were obscured by scattered light. As such, Sector 27 was not included in our analysis. Table \ref{tab:all_times} shows all derived transit midpoint posteriors. Table \ref{tab:RV_datasets} details all radial velocity data sources while all RV data points can be found in Table \ref{tab:all_RV}. All TESS photometry and radial velocity data used in this research are open source and available to the public for further perusal.

\begin{figure*}[p]
  \begin{center}
    \includegraphics[width=\textwidth]{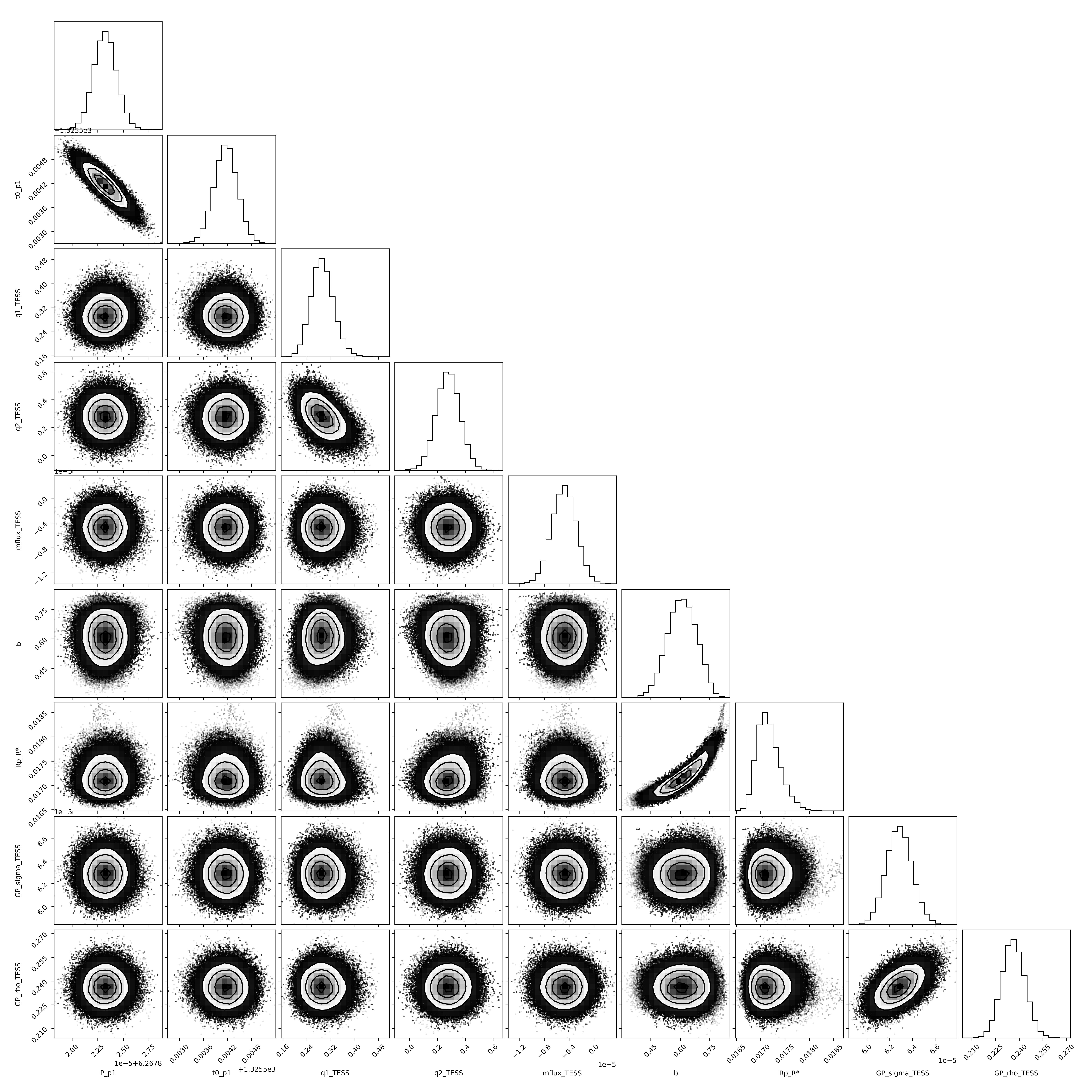} \\
    \caption{Corner plot of $\pi$~Men~c planetary parameter posteriors as derived by the photometry TTV \texttt{juliet} nested sampling model. All distributions are consistent with Gaussian distributions.}
    
    \label{fig:corner_phot}
  \end{center}
\end{figure*}

\begin{figure*}[p]
  \begin{center}
    \includegraphics[width=\textwidth]{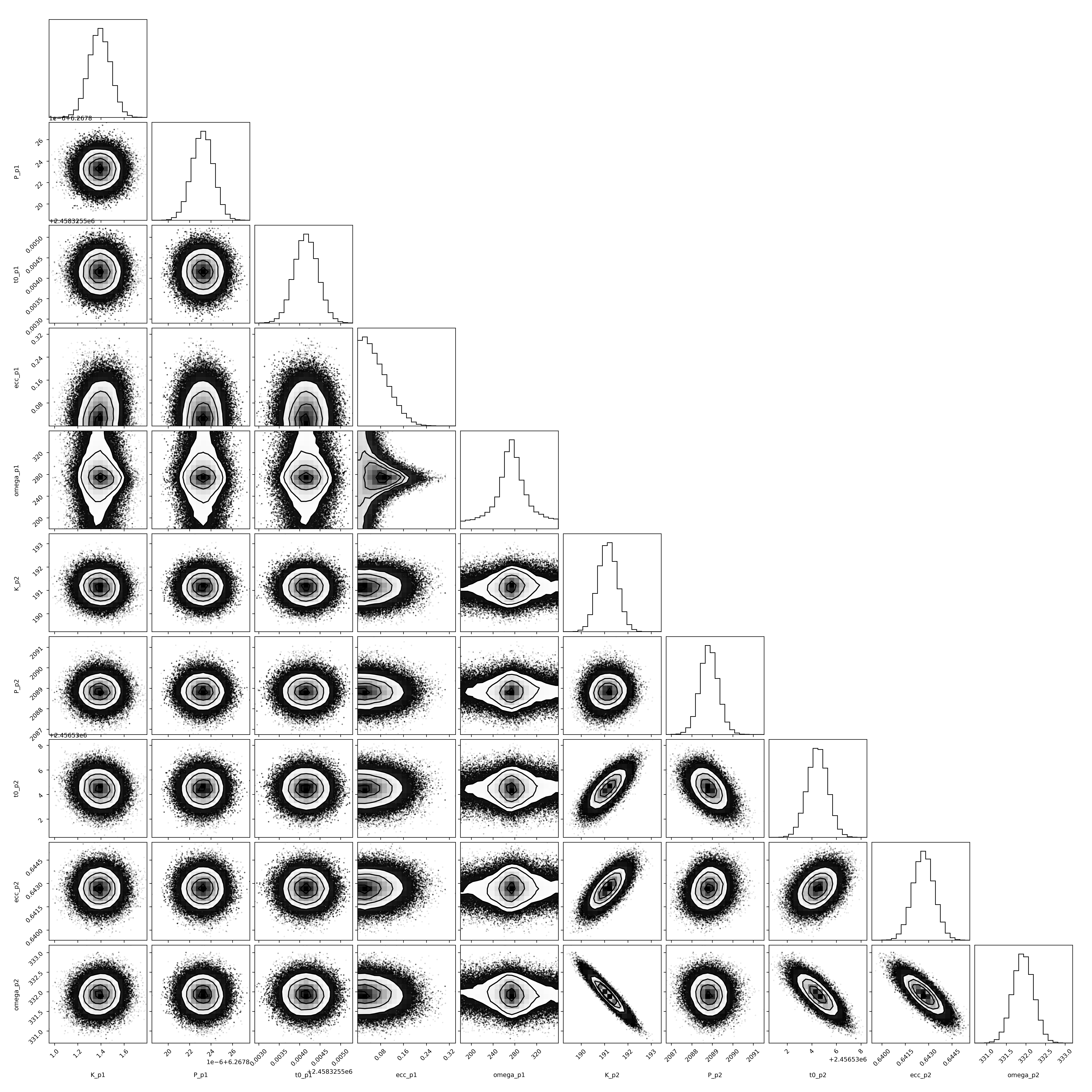} \\
    \caption{Corner plot of $\pi$~Men~c (labeled as p1) and b (labeled as p2) as planetary parameter posteriors as derived by the 2-planet radial velocity \texttt{juliet} nested sampling model. Note that "planet order" (p1, p2, p3) is designated by period with the innermost planet listed first, not by discovery date.
    }
    \label{fig:corner_RV_2planets}
  \end{center}
\end{figure*}

\begin{figure*}
  \begin{center}
    \includegraphics[width=\textwidth]{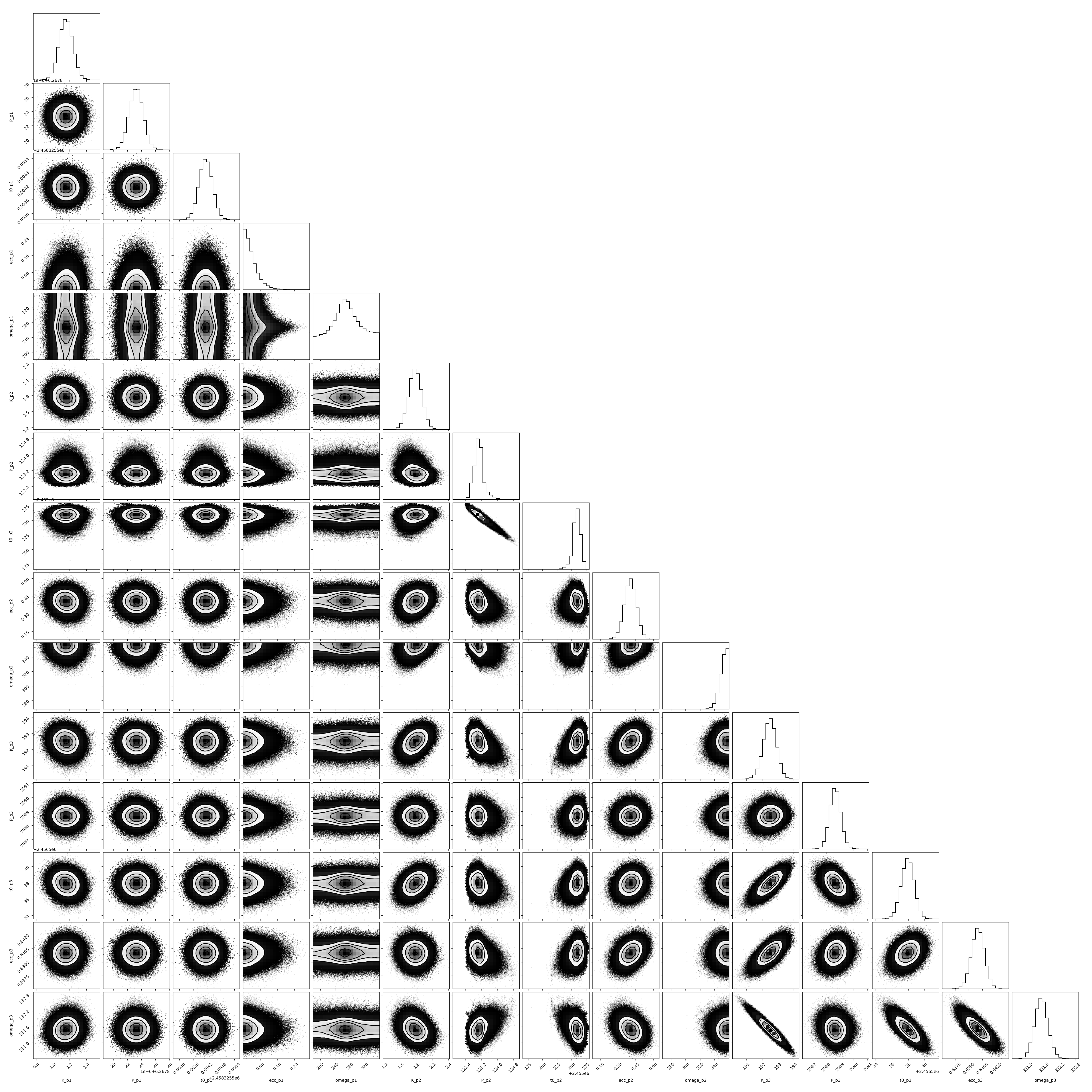} \\ 
    \caption{Corner plot of $\pi$~Men~c (p1), d (p2), and b (p3) planetary parameter posteriors as derived by the 3-planet radial velocity \texttt{juliet} nested sampling model. Note that "planet order" (p1, p2, p3) is designated by period with the innermost planet listed first, not by discovery date.
    }
    \label{fig:corner_RV_3planets}
  \end{center}
\end{figure*}

Figure \ref{fig:corner_phot} shows the corner plot of $\pi$~Men~c planetary parameter distributions via TESS photometry TTV model. Figures \ref{fig:corner_RV_2planets} and \ref{fig:corner_RV_3planets} show the corner plots of the $\pi$ Men planetary parameters analyzed in the RV 2-planet and 3-planet models respectively. Planets are labeled in order of period size; for example, $\pi$~Men~c is listed as p1 because it consistently has the innermost period regardless of model.

\begin{table*}[tbhp!]
\caption{TESS Photometry Sectors}
    \centering
    \begin{tabular}{lcc}
    \hline
    Sector  & $\#$ Transits & Timeframe (YYYY MM DD) \\
    \hline
    s001   & 5  &  2018 07 25 - 2018 08 22  \\
    s004   & 2*  &  2018 10 19 - 2018 11 14  \\
    s008   & 3  &  2019 02 02 - 2019 02 27  \\
    s011   & 4  &  2019 04 23 - 2019 05 20  \\
    s012   & 4  &  2019 05 21 - 2019 06 18  \\
    s013   & 4  &  2019 06 19 - 2019 07 17  \\
    s028   & 4  &  2020 07 31 - 2020 08 25  \\
    s031   & 4  &  2020 10 22 - 2020 11 18  \\
    s034   & 4  &  2021 01 14 - 2021 02 08  \\
    s038   & 4  &  2021 04 29 - 2021 05 26  \\
    s039   & 4  &  2021 05 27 - 2021 06 24  \\
    s061   & 4  &  2023 01 18 - 2023 02 12 \\
    s062   & 4  &  2023 02 12 - 2023 03 10  \\
    s064   & 4  &  2023 04 06 - 2023 05 04  \\
    s065   & 4  &  2023 05 04 - 2023 06 02  \\
    s066   & 2  &  2023 06 02 - 2023 07 01  \\
    s067   & 3  &  2023 07 01 - 2023 07 29  \\
    s068   & 2**  &  2023 07 29 - 2023 08 25 \\
    s088   & 3  &  2025 01 14 - 2025 02 11  \\
    s089   & 5  &  2025 02 11 - 2025 03 12  \\
    s093   & 4  &  2025 06 03 - 2025 06 29  \\
    s094   & 4  &  2025 06 29 - 2025 07 25  \\
    s095   & 2  &  2025 07 25 - 2025 08 20  \\
    \hline
    \end{tabular}
    \label{tab:Phot_sectors}
\tablecomments{*There are 3 visible $\pi$~Men~c transits in Sector 4, but one transit was excluded for being obstructed by turbulent stellar noise.}
\tablecomments{**There are 3 visible $\pi$~Men~c transits in Sector 68, but one transit was excluded for being obstructed by turbulent stellar noise.}
\tablerefs{doi:\href{https://archive.stsci.edu/doi/resolve/resolve.html?doi=10.17909/wsmt-mv18}{10.17909/wsmt-mv18}}
\end{table*}

\begin{table*}[tbhp!]
    \caption{All TESS transit times derived in this paper}
    \centering
    \begin{tabular}{cccc}
    \hline
       Mid-transit time     & Upper Error & Lower Error & Epoch  \\
     (BTJD; BJD - 2457000)   & (days)  & (days) & \\
    \hline
1325.50416 &	0.00028	&	0.00029 &	0 \\
1331.7704  &    0.0020  &   0.0025  & 1 \\
1338.0391  &    0.0014  &   0.0016  & 2 \\
... & ... & ... & ... \\
3876.5088  &    0.0010  &   0.0011  & 407 \\
3889.0444  &    0.0010  &   0.0009  & 409 \\
3901.5785  &    0.0012  &   0.0012  & 411 \\

    \hline     
    \end{tabular}
    \label{tab:all_times}
    \tablecomments{
    This table is available in its entirety in machine-readable form.   
    }
\end{table*}

\begin{table*}[p]
\caption{Radial Velocity Datasets}
    \centering
    \begin{tabular}{lccccc}
    \hline
    Instrument & $N_{meas.}$ & Timeframe  & Timeframe & RV Source & Original Source\\
                &            &    (BJD-2450000) &      (YYYY MM DD)  \\                 
    \hline
    UCLES        & 42 & 830.0 -- 3669.2   &  1998 01 16 -- 2005 10 25  &  [1] & [2]; [3] \\
    CORALIE-98   & 10  & 1139.8 -- 4108.7  &  1998 11 22 -- 2007 01 08  & [4]  &   [4]  \\
    CORALIE-07   & 12  & 4433.7 -- 6586.8  &  2007 11 29 -- 2013 10 21  & [4]  &   [4]  \\
    CORALIE-14   & 38  &  6648.8 -- 8891.5  &  2013 12 22 -- 2020 02 12  & [4]  &  [4]  \\
    ESPRESSO     & 275 &  8367.9 -- 8568.5  &  2018 09 06 -- 2019 03 26  & [4]  &  [4] \\
    HARPS\tablenotemark{$*$} & 402 & 8383.9 -- 9185.7  &  2018 09 22 -- 2020 12 02  & [5]  &  [5]   \\

    \hline
    \end{tabular}
    \label{tab:RV_datasets}
\tablecomments{*Listed in \cite{Hatzes_2022} as "HARPS-POST (Large)"}
\tablerefs{
[1] \cite{Gandolfi_2018}; 
[2] \cite{Jones_2002}; 
[3] \cite{Butler_2006};
[4] \cite{Damasso_2020};
[5]\cite{Hatzes_2022}
}
\end{table*}

\begin{table}[tbhp!]
    \caption{All $\pi$ Men Radial Velocity Measurements}
    \centering
    \begin{tabular}{ccccc}
    \hline
    Time          & RV          & $\sigma_{RV}$ & Instrument  & RV Source \\
    (BJD$_{TDB}$) & (m s$^{-1}$)  & (m s$^{-1}$)\\
    \hline
    2450829.993723 & -41.0  & 4.8         & UCLES          &  [1] \\
    2451119.251098 & -67.4  & 9.8 & UCLES & [1] \\
    2451236.033635 & -79.2  & 6.0 & UCLES & [1] \\ 
    ... & ... & ... & ... & ... \\
    2458844.690365 & 10724.36 & 3.23 & CORALIE-14 & [2] \\
    2458855.703833 & 10715.16 & 4.42 & CORALIE-14 & [2] \\
    2458891.543187 & 10707.57 & 2.94 & CORALIE-14 & [2] \\
    \\
    \hline     
    \end{tabular}
    \label{tab:all_RV}
    \tablecomments{
    This table is available in its entirety in machine-readable form.
    \tablerefs{
    1.) \citet{Gandolfi_2018}, 2.) \citet{Damasso_2020}, 3.) \citet{Hatzes_2022}
    }}
\end{table}

\newpage 
\bibliography{Pi_men_c_bib}{}
\bibliographystyle{aasjournalv7}

\end{document}